\documentclass[12pt]{article}
\pdfoutput=1
\usepackage{hyperref}
\usepackage{booktabs}
\usepackage{multirow}
\usepackage{dcolumn}
\usepackage{gensymb}
\usepackage{textgreek}
\usepackage[detect-all]{siunitx}
\usepackage{soul}

\usepackage[a4paper, total={6in, 8in}]{geometry}
\usepackage[super]{natbib}

\usepackage{setspace}
\counterwithin{figure}{section}

\AtBeginDocument{
\heavyrulewidth=.08em
\lightrulewidth=.05em
\cmidrulewidth=.03em
\belowrulesep=.65ex
\belowbottomsep=0pt
\aboverulesep=.4ex
\abovetopsep=0pt
\cmidrulesep=\doublerulesep
\cmidrulekern=.5em
\defaultaddspace=.5em
}

\newcolumntype{P}[1]{>{\centering\arraybackslash}p{#1}}

\newcommand{\lsim}{\mathrel{\mathop{\kern 0pt \rlap
  {\raise.2ex\hbox{$<$}}}
  \lower.9ex\hbox{\kern-.190em $\sim$}}}
\newcommand{\gsim}{\mathrel{\mathop{\kern 0pt \rlap
  {\raise.2ex\hbox{$>$}}}
  \lower.9ex\hbox{\kern-.190em $\sim$}}}

\usepackage{amsmath}
\usepackage{graphicx,bm}

\usepackage[caption=false]{subfig}
\usepackage{slashed}
\usepackage{setspace}
\usepackage[hang]{footmisc}
\usepackage{lipsum}
\usepackage{listings}

\usepackage[usenames,dvipsnames]{color}

\usepackage[utf8]{inputenc}

\definecolor{MyDarkGreen}{rgb}{0.0,0.4,0.0}

\newcommand{\red}{\textcolor{red}}

\newcommand{\Hmat}{\boldsymbol{H}}
\newcommand{\invH}{\boldsymbol{H^{-1}}}

\usepackage{authblk}

\begin{document}

\title{The hybrid approach -- Convolutional Neural Networks and Expectation Maximization Algorithm -- for Tomographic Reconstruction of Hyperspectral Images}

\date{}

\author[,1]{\small{Mads Juul Ahleb{\ae}k}\footnote{ahle@sdu.dk}}
\author[,2,3]{\small{Mads Svanborg Peters}\footnote{mape@newtec.dk}}
\author[,1]{\small{Wei-Chih Huang}\footnote{huang@cp3.sdu.dk}}
\author[,1]{\\ \small{Mads Toudal Frandsen}\footnote{frandsen@cp3.sdu.dk}}
\author[,3]{\small{Ren\'e Lynge Eriksen}\footnote{rle@mci.sdu.dk}}
\author[,2]{\small{Bjarke J{\o}rgensen}\footnote{bjarke@newtec.dk}}

\affil[1]{\small{CP$^3$-Origins, Department of Physics, Chemistry and Pharmacy, University of Southern Denmark, Denmark}}
\affil[2]{\small{Newtec Engineering A/S, 5230 Odense, Denmark}}
\affil[3]{\small{Mads Clausen Institute, University of Southern Denmark, Denmark}}







\maketitle

\begin{abstract}
We present a simple but novel hybrid approach to hyperspectral data cube reconstruction from computed tomography imaging spectrometry~(CTIS) images that sequentially combines neural networks and the iterative Expectation Maximization~(EM) algorithm.
We train and test the ability of the method
to reconstruct data cubes of $100\times100\times25$ and $100\times100\times100$ voxels, corresponding to 25 and 100 spectral channels,
from simulated CTIS images generated by our CTIS simulator.
The hybrid approach utilizes the inherent strength of the Convolutional Neural Network~(CNN) with regard to noise and its ability to yield consistent reconstructions
and make use of the EM algorithm's ability to generalize to spectral images of any object without {\it training}.

The hybrid approach achieves better performance than both the CNNs and EM alone for seen~(included in CNN training) and unseen~(excluded from CNN training)  cubes for both the 25- and 100-channel cases. For the 25 spectral channels, the improvements from CNN to the hybrid model~(CNN + EM) in terms of the mean-squared errors are between $14 - 26 \ \%$.
For 100 spectral channels, the improvements between
$19-40 \ \%$ are attained with the largest improvement of $40 \ \%$ for the unseen data, to which the CNNs are not exposed during the training.
\end{abstract}


\section*{Keywords}
Snapshot, Hyperspectral imaging, Artificial Neural Networks, Convolutional Neural Networks, Tomographic reconstruction

\section{Introduction \label{sec:introduction}}

Multispectral and Hyperspectral Imaging~(MSI, HSI)~\cite{Goetz1147} is used in a wide range of applications in diverse fields. These include astronomy and space surveillance~\cite{article}, spectroscopic differentiation of materials in geoscience~\cite{keshava_distance_2004}, detection of foreign objects and weeds in precision agriculture~\cite{lee_non-destructive_2017}, and optical sorting within the food industry~\cite{pu_recent_2015}. 
HSI produces 3-dimensional~(3-D) data cubes that capture light intensities in two spatial and one spectral dimension.
Pushbroom~(line scan)~\cite{boldrini_hyperspectral_2012} HSI is the standard technique but it requires steady movement of either the object or camera to acquire a hyperspectral image. Also, the equipment cost is typically high, and this creates barriers to broader applications of HSI.

On the other hand, the Computed Tomography Imaging Spectrometer~(CTIS)~\cite{Okamoto:91,Th,descour_computed-tomography_1995}  is a relatively simple, and potentially compact and cheap snapshot HSI system, which can capture an image within milliseconds or shorter.
There exist alternative snapshot spectral imaging technologies that capture (projections of) the 3-D data cube instantaneously using dispersive optics such as
single-shot compressive spectral imaging with a dual-disperser architecture~(CASSI)~\cite{gehm_single-shot_2007}, 
Hybrid camera Multispectral-Video Imaging System (HMVIS)~\cite{cao_high_2011}, 
lenslet-array~\cite{bodkin_snapshot_2009},
filter-on-chip imagers \cite{von_freymann_compact_2014},
Image Mapping Spectrometers (IMS) \cite{Gao:10}, Image-replicating Imaging Spectrometers \cite{10.1117/12.580059} and snapshot HSI Fourier transform spectrometer~\cite{Kudenov:12}. Nonetheless, the CTIS is investigated here.

\begin{figure}[htp]
\begin{center}
\includegraphics[width=0.8\textwidth]{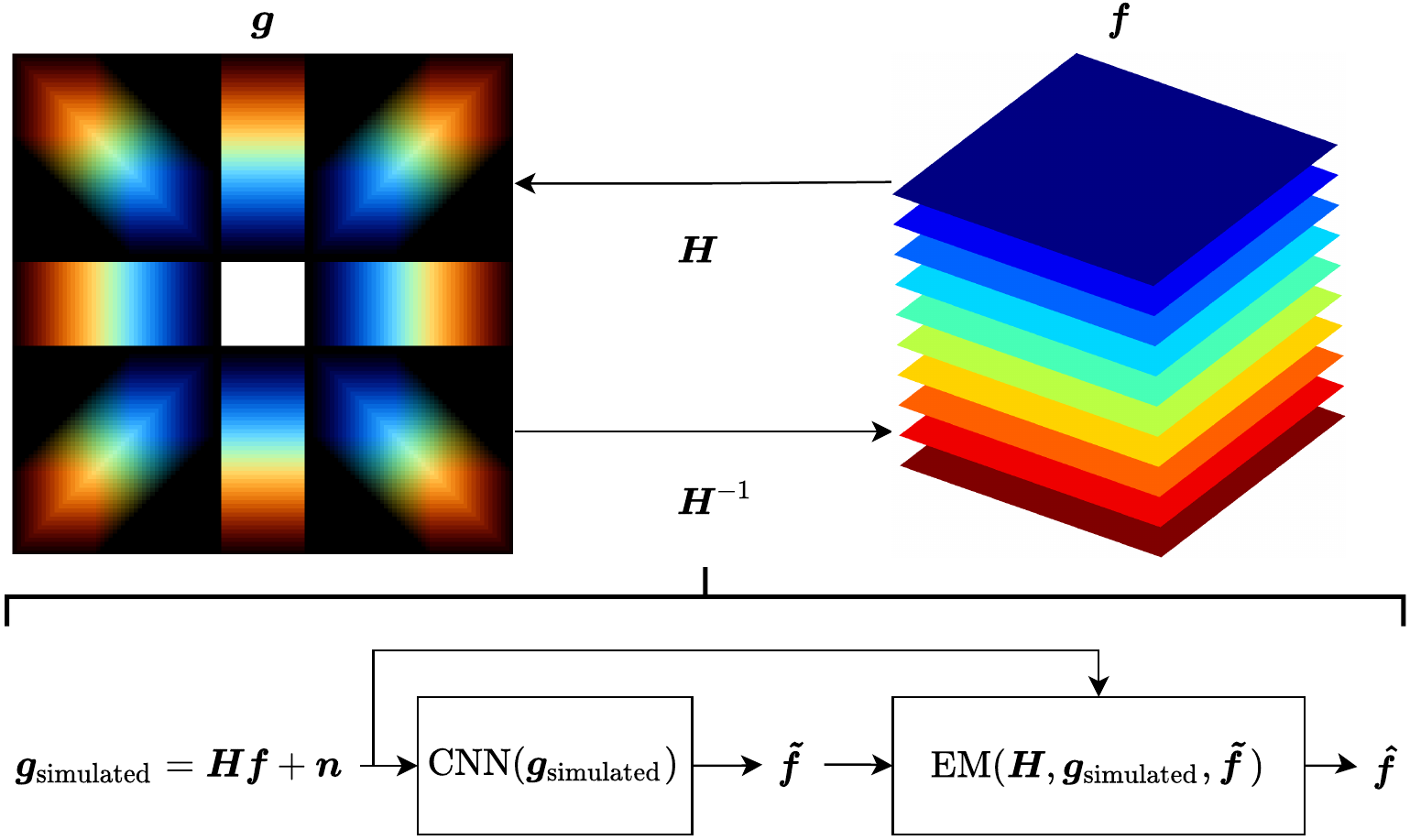}
\caption{Top: Illustration of the connection between a hyperspectral cube $\boldsymbol{f}$ and the corresponding CTIS image $\boldsymbol{g}$ via the $\Hmat$ matrix. The image $\boldsymbol{g}$ consists of a central zeroth diffraction order and eight surrounding first diffraction orders, which are projections of the hyperspectral cube $\boldsymbol{f}$.
Bottom: The workflow of the hybrid approach, where $\boldsymbol{\tilde{f}}$ and $\boldsymbol{\hat{f}}$ are estimates of $\boldsymbol{f}$ reconstructed with the CNN and hybrid (CNN+EM), respectively; see explanations in the main text.}
\label{Fig:hybrid}
\end{center}
\end{figure}

The CTIS system acquires a 2-D image $\boldsymbol{g}$ by means of a diffractive optical element (DOE) that diffracts the 3-D hyperspectral cube $\boldsymbol{f}$ into the zeroth and surrounding first orders (Figure~\ref{Fig:hybrid}), corresponding to projections of the cube onto a 2-D plane.
The 2-D projection is determined by the system matrix, denoted by
$\Hmat$~($\boldsymbol{g} = \Hmat \boldsymbol{f}$), which incorporates the optical parameters of the CTIS system as we shall discuss below. 

For a captured CTIS image $\boldsymbol{g}$, one can either directly analyze it, e.g., for classification of apple scab lesions~\cite{Douarre:20,douarre_ctis-net_2021}, or reconstruct  a 3-D cube $\boldsymbol{f}$ from $\boldsymbol{g}$, which leads to wider applications, with the help of the inverse matrix $\invH$.
However, $\Hmat$ is a sparse, enormous, and (usually) rectangular matrix, which makes computation of the Moore-Penrose pseudo-inverse impractical in terms of both computation time and memory consumption. 
As a result, iterative algorithms~\cite{vose_heuristic_2007,hagen_fourier_2007,white_accelerating_2020} have been proposed to reconstruct $\boldsymbol{f}$.
The reconstruction time is unfortunately quite long and the accuracy is mediocre, especially for large images or high numbers of spectral channels, which hinders practical applications. 
Fast and precise real-time reconstruction is therefore an important but challenging goal.

In Ref.~\cite{huang2022application}, we, for the first time, applied Convolutional Neural Networks~(CNN)~\cite{6795724, 726791} to reconstruct the data cube from a simulated CTIS image.
By comparison, we also used the Expectation Maximization (EM) algorithm for cube reconstruction.
The EM takes as input arguments $\Hmat$, a CTIS image $\boldsymbol{g}$ and an initial guess of the reconstructed 3-D cube,
and iteratively updates the cube until it approximately reproduces the true cube. 
Overall, the CNN performance is much better with a shorter reconstruction time than the EM.
Moreover, the network can handle images of different objects, yielding a consistent accuracy whereas
the EM is challenged with objects of complex geometry~(complex geometries having high spatial frequencies and varied spectral information).
The EM algorithm is also susceptible to inherent noise in the CTIS images and fails to converge for noisy CTIS images. That is, the difference between the true and EM-reconstructed cube increases with more iterations \cite{snyder_noise_1987,scholl_design_2010,zeng_unmatched_2000}.
An important caveat is that the CNN has to see similar objects or geometries in the training phase to make reliable reconstructions.
In other words, it cannot manage objects very different from those in the training data, whereas the EM applies to images of any objects.

In this work, we propose a simple but novel hybrid method that sequentially combines the neural networks and EM to circumvent the
aforementioned shortcomings intrinsic to the EM algorithm and networks, respectively. Thus, a network is first employed to reconstruct a data cube from a CTIS image and then its output, as an initial guess  of the data cube, is passed to the EM algorithm, which further improves the
network predictability as depicted in Figure~\ref{Fig:hybrid}.
In other words, the network provides a refined initial condition for the EM algorithm that is close to the correct data cube to guarantee convergence.
Furthermore, the existence of the EM algorithm as the second step ensures that the hybrid model can be applied to new types of images even if those are not included in the training data of the network.
%


\section{Introduction to CTIS imaging system and procedures of data generation}\label{Sec:CTIS_sim}

In this section, we introduce the CTIS imaging system, discuss the updated CTIS simulator, and detail the preparation of data for 25 and 100 spectral channels used in the training and testing of the neural networks.
\subsection{CTIS imaging system}

The CTIS imaging system can be described by the linear imaging
equation~\cite{descour_computed-tomography_1995}:
\begin{align}\label{eq:gHf}
	\boldsymbol{g} = \boldsymbol{H}\boldsymbol{f} + \boldsymbol{n}
\end{align}
where $\boldsymbol{g}$ is a vectorized CTIS image vector of $q^2$ elements and $\boldsymbol{f}$ is the vectorized hyperspectral cube with $r=x \cdot y \cdot z$ voxels, where $x,y$ and $z$ denote the two spatial dimensions and the number of spectral channels, respectively,
while $\boldsymbol{n}$ corresponds to a random noise vector.
The $q^2\times r$ system matrix $\boldsymbol{H}$ describes the projection of the \textit{i}-th voxel in $\boldsymbol{f}$ to the \textit{j}-th pixels in $\boldsymbol{g}$ - equivalent to the nine projections in Figure~\ref{Fig:hybrid}, which consist of a central zeroth order, surrounded by eight first orders.

The system matrix $\boldsymbol{H}$ is constructed assuming spatial shift-invariance and a linear mapping between $\boldsymbol{f}$ and $\boldsymbol{g}$. It includes the point spread function (PSF), the illumination (wavelength-dependent intensity) and the diffraction sensitivity (which includes diffraction efficiency of the DOE, the transmission of the optical system and the sensor response). Both the illumination and diffraction sensitivity are wavelength-dependent, while the PSF is assumed wavelength-independent in our limited wavelength range and with the used optics. Additionally, the diffraction sensitivity depends on the respective zeroth or first orders.
See Supplementary Material Section~S3-S5 for additional details on determining system parameters. Thus, the \textit{i}-th voxel, $f_i$, is mapped into each diffraction order with a sensitivity given by the product of the diffraction sensitivity and the illumination for a specific wavelength and diffraction order. The resulting $q\times q$ CTIS image is convolved with the PSF, vectorized and arranged as columns in $\boldsymbol{H}$ for $i = 1,\dots,r$. Due to the sparsity of $\boldsymbol{H}$, it is naturally implemented as a sparse matrix, which significantly decreases both the computation time and memory consumption.
But even with a sparse $\Hmat$, the memory requirements are large for large data cube dimensions ($x,y\geq 100 \ \text{and} \ z\geq 25$). Since a significant amount of CTIS images are needed for the training of the neural networks, a CTIS simulator was used to generate images without resorting to a system matrix $\Hmat$.

\subsection{Simulating CTIS images for 25 and 100 spectral channels}

Our updated CTIS simulator was created with generalizability in mind.
It generates a CTIS image from an input hyperspectral cube which can have arbitrary spatial and spectral dimensions while enabling control of both geometric and optical parameters.
The main purpose of the simulator is to speed up the generation of CTIS images for the training of the neural networks and remove the need for a system matrix for large cube dimensions.

We have significantly improved the CTIS simulator employed in our previous work~\cite{huang2022application}: The updated simulator executes Eq.~\eqref{eq:gHf}, and similarly to the system matrix incorporates a PSF, spectral sensitivity corrections in terms of diffraction sensitivity and illumination as well as additive zero-mean Gaussian noise.
It emulates our laboratory CTIS system~\cite{peters_high-resolution_2022}, from which the optical parameters used in the simulator have been measured; 
see Supplementary Material Section~S2-S5 for more details. The scripts for the CTIS simulator and generation of $\Hmat$ in \texttt{MATLAB} and \texttt{Python} are available on \href{https://github.com/madspeters/CTIS}{Github}. 
\begin{figure}[htp]
	\begin{center}
		\includegraphics[width=\textwidth]{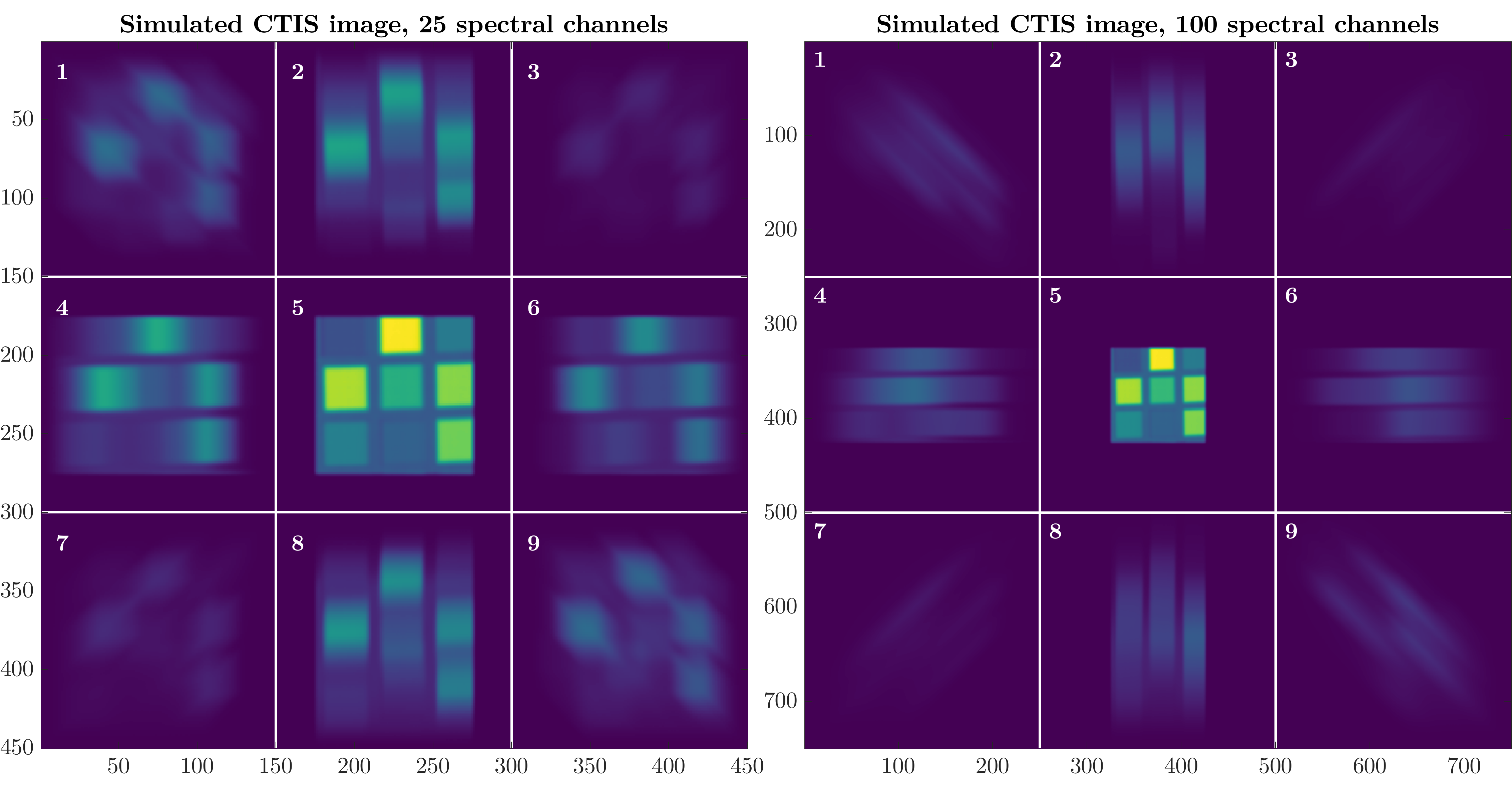}
		\caption{Left (Right): A simulated $450\times450$ ($750\times750$) pixels CTIS image for a $100\times100\times25$ ($100\times100\times100$) data cube of a Colorchecker divided into 9 smaller $150\times150$ ($250\times250$) pixels images, which are fed into the CNNs.}
		\label{Fig:CTISSimulator}
	\end{center}
\end{figure}

A simulated $450\times450$ pixels CTIS image generated from a $100\times100\times25$ data cube is shown on the left in Figure~\ref{Fig:CTISSimulator}. All simulated CTIS images comprise the central zeroth order and eight neighboring first orders.  The geometric parameters of the simulator enable control of the cube dimensions, the distance between the zeroth and first orders and the pixel shift between projections of the spectral channels in the first orders. In Figure~\ref{Fig:CTISSimulator} the geometric parameters are a shift of 2 pixels, a 27-pixel distance between the zeroth and first order and the $100\times100\times25$ dimensions of the data cube. 
Note that the simulated images from the chosen parameters are smaller than those captured by our CTIS camera.
Additionally, the determined optical parameters are also incorporated as seen from the nonuniform intensities among  the first orders. 
To assist the networks in identifying regions of interest on CTIS images, we pre-process the 2-D CTIS images with the division into 9 smaller $150\times150$ images, each containing either the zeroth or first order as shown in the left panel of Figure~\ref{Fig:CTISSimulator}.
For the 100-channel case (right panel of Figure~\ref{Fig:CTISSimulator}), the optical parameters are updated correspondingly to match the higher number of spectral channels. That results in a $750 \times 750$ pixels CTIS image, which is divided into 9 smaller $250 \times 250$ images as in the 25-channel case.

As our main goal is to reconstruct 3-D hyperspectral cubes from real CTIS image, the data cubes used to simulate CTIS images (and used as ground truth in training) are captured by our pushbroom HSI system. The system consists of a conveyor belt and an HSI camera, which contains an ImSpector V10E spectrograph (Specim), a 50~mm C Series VIS-NIR objective (Edmund Optics) and a Qtechnology QT5022 system equipped with a CMV4000-E12 CMOS sensor (CMOSIS). The pushbroom system acquires 216 spectral channels between the wavelengths 384 and 972~nm with a spatial resolution of 0.33~mm pixels$^{-1}$.

\subsection{Data preparation}
The data used in this work originate from  178  different pushbroom cubes of various objects, such as potatoes, a Colorchecker and books, with varying spatial dimensions, ranging from $200\times400$ to $499\times400$, all with 216 channels.
Seven of these cubes are reserved as completely unseen cubes for testing the networks' capability of generalization, while the remaining 171 cubes are used for training, validation and testing. These unseen cubes contain pears, potatoes with wireworm defects and carrots.
\begin{figure}[h]
	\centering
	\includegraphics[width=\textwidth]{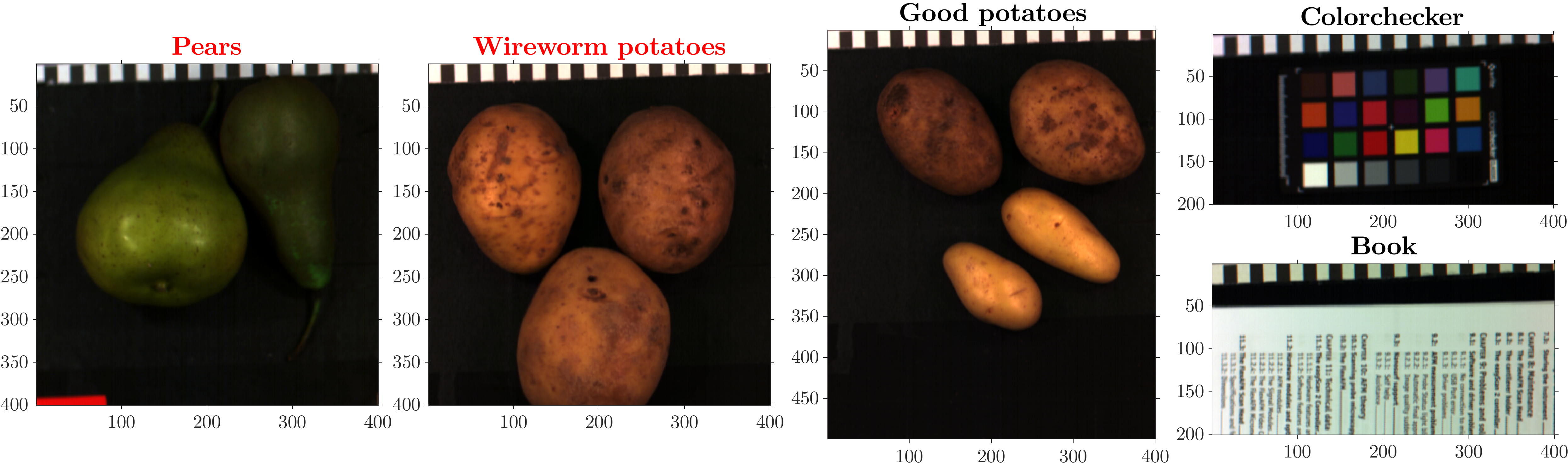}
	\caption{RGB visualization examples of seen (unseen) hyperspectral cubes with black (red) titles captured by the pushbroom system.
	The examples of unseen cubes consist of pears and potatoes with wireworms, while the seen cubes consist of good potatoes, a Colorchecker and a book. RGB images are created by combining the 470~nm (blue), 549~nm (green) and 650~nm (red) spectral channels.}
	\label{Fig:cube_100_demon}
\end{figure}
RGB visualizations of some of the used pushbroom cubes created by combining three spectral channels at 470~nm (blue), 549~nm (green) and 650~nm (red) are displayed in Figure~\ref{Fig:cube_100_demon}. RGB images of all 178 cubes are presented in Supplementary Material in Section~S6. 

\begin{figure}[h]
	\centering
	\includegraphics[width=\textwidth]{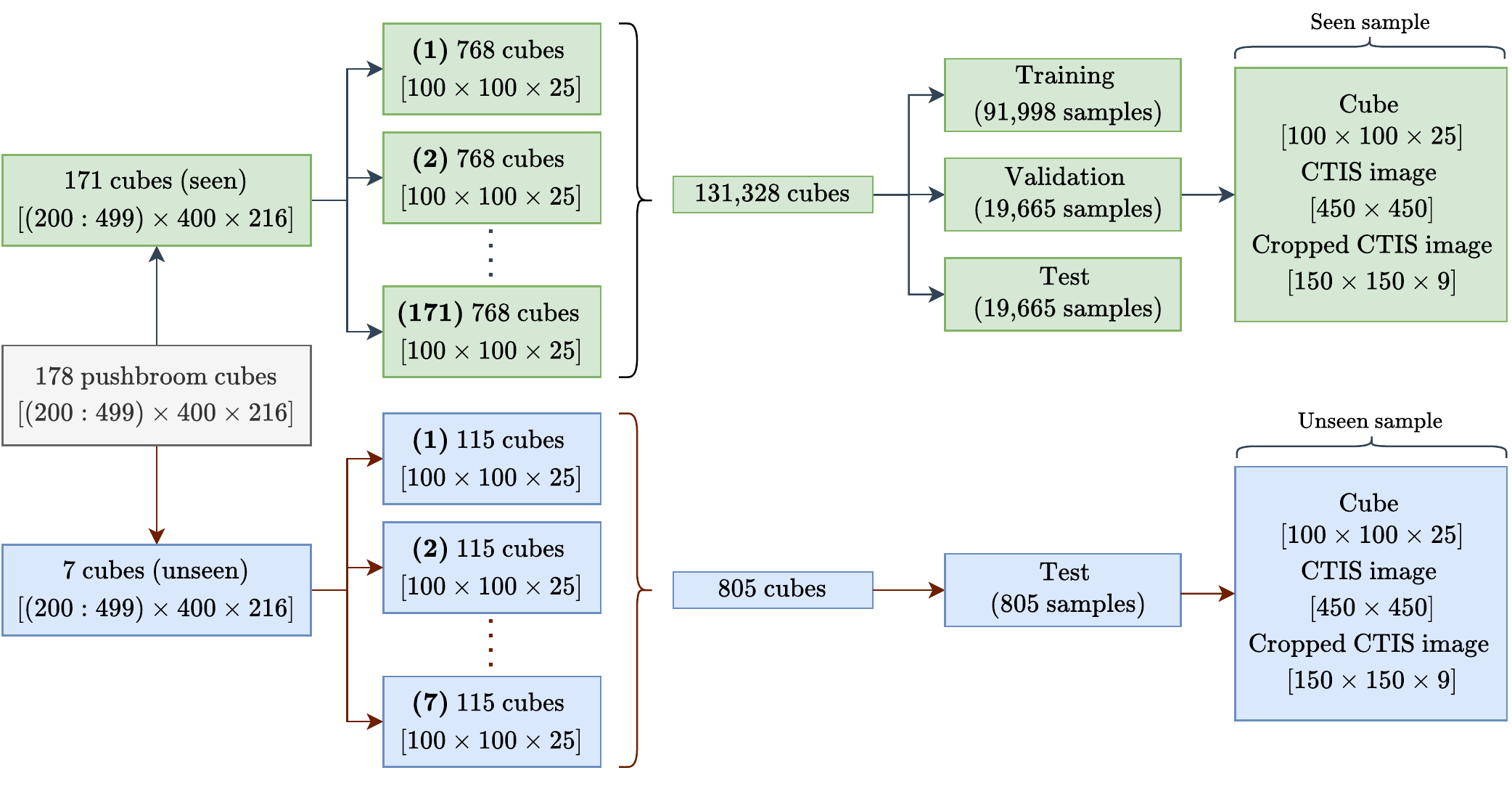}
	\caption{Overview of the data pipeline for the data generation with 25 spectral channels: The captured 178 pushbroom cubes are grouped into 171 seen and 7 unseen cubes, which are then cropped to smaller $100\times100\times25$ cubes. The samples are  divided into training, validation and test sets for the neural networks.}
	\label{Fig:dataGeneration}
\end{figure}
The data preparation for the neural networks is illustrated in Figure~\ref{Fig:dataGeneration}: For each of the seen 171 pushbroom cubes, we crop 768 smaller cubes of dimension (100, 100, 25). The 216 spectral channels are reduced to 25 by removing the first 10 and last 6 spectral channels, which have a low signal-to-noise ratio, and averaging over 8 consecutive spectral channels for the remaining 200 spectral channels - resulting in 25 channels. Then, the simulator is applied to these smaller cubes to generate CTIS images. 
In total, there are 131,328 samples, which are divided into training~(91,998), validation~(19,665) and test~(19,665) sets.
The training set is used to train networks, while the validation set is employed to prevent overfitting.
We evaluate the models via the test set that has not been involved in the training process.  
Besides, we create 805 extra samples from the 7 unseen cubes to assess how well the models can handle completely new data.
All in all, each sample contains a  $100\times100\times25$ cube, a full $450\times450$ CTIS image and the corresponding pre-processed $150\times150\times9$ CTIS image.  
The full CTIS images are used in the EM algorithm, while the neural networks take the pre-processed CTIS images as input and hyperspectral cubes as output. For the 100-channel case, the procedure of data generation is the same, where a sample contains a $100\times100\times100$ data cube, a full $750\times750$ pixels CTIS image and the corresponding pre-processed $250\times250\times9$ CTIS image.

\section{Models of data cube reconstruction with 25 spectral channels from CTIS images}\label{Sec:25channel}
In this section, we elaborate on three different methods of hyperspectral data cube reconstruction from CTIS images: the EM algorithm, CNNs and hybrid CNN-EM models.
The models will be trained and tested on data consisting of $450\times 450$ pixels CTIS images cropped into 9 regions of $150\times 150$ pixels as input
and hyperspectral cubes of $100\times100$ pixels with 25 spectral channels as output.
\subsection{EM reconstruction algorithm}
The EM algorithm~\cite{shepp_maximum_1982} is routinely utilized in the CTIS reconstruction~\cite{descour_computed-tomography_1995,wilson_reconstructions_1997,hagen_fourier_2007}.
Since $\boldsymbol{H}$ is generally non-invertible, an estimate  $\boldsymbol{\hat{f}}$ of the hyperspectral cube  is obtained using a sparse implementation of the iterative EM algorithm, which effectively attempts to solve Eq.~\eqref{eq:gHf} for a given CTIS image $\boldsymbol{g}$. The EM algorithm first computes an estimated CTIS image $\boldsymbol{\hat{g}} = \boldsymbol{H} \boldsymbol{\hat{f}}^{(k)}$ in the expectation step. The subsequent maximization step computes a correction factor for all voxels in $\boldsymbol{\hat{f}}^{(k)}$ as a back-projection of the ratio of the acquired $\boldsymbol{g}$ and estimated CTIS image $\boldsymbol{\hat{g}}$, normalized by the summed rows of $\boldsymbol{H}$:

\begin{align}\label{eq:EM_1}
	\boldsymbol{\hat{f}}^{(k+1)}	 = \frac{\boldsymbol{\hat{f}}^{(k)}}{\sum_{i=1}^{q^2} H_{ij}}\odot \left( \boldsymbol{H}^T \frac{\boldsymbol{g}}{\boldsymbol{H}\boldsymbol{\hat{f}}^{(k)}}\right)
\end{align}
where $k$ is the iteration index, $\boldsymbol{\hat{f}}^{(k)}$ is the $k$-th estimate of the hyperspectral cube, $\sum_{i=1}^{q^2} H_{ij}$ is the vectorized summation of rows in $\boldsymbol{H}$, $\boldsymbol{H}^T$ is the transposed system matrix, and the symbol $\odot$
denotes the Hadamard~(elementwise) product. Notice that Eq.~\eqref{eq:EM_1} combines the  expectation and maximization steps into a single step.

Initialization is typically either $\boldsymbol{\hat{f}}^{(0)} = ones(r,1)$~\cite{wilson_reconstructions_1997} or $\boldsymbol{\hat{f}}^{(0)} = \boldsymbol{H}^T\boldsymbol{g}$~\cite{descour_computed-tomography_1995}, where the former  is utilized in this work for EM reconstructions. As 10-30 EM iterations are typically required~\cite{white_accelerating_2020},
we chose to use 20 iterations for the standalone EM and apply only 10 iterations for the hybrid models.

\subsection{Convolutional Neural Network}

\begin{figure}[htp]
	\begin{center}
		\includegraphics[width=0.9\textwidth]{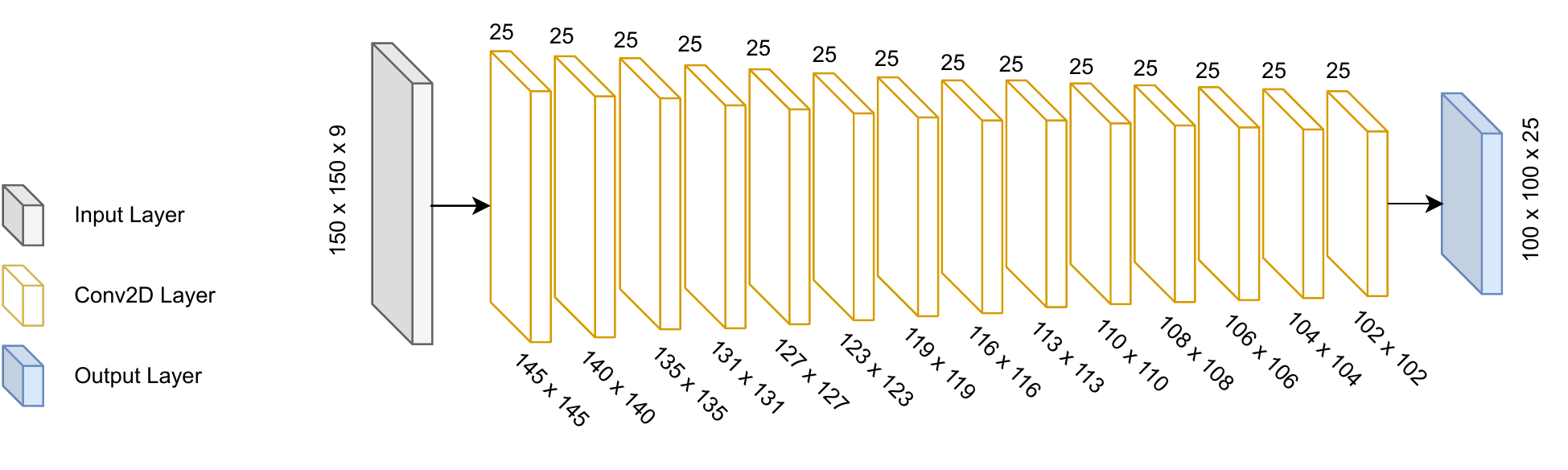}
		\caption{Illustration of the proposed CNN1 for reconstructing a $100\times100$ image with 25 spectral channels. The network has a total of 174,100 trainable parameters and consists of an input layer~(which takes CTIS images), \texttt{Conv2D} layers and an output layer~(which generates hyperspectral cubes).
		For the hidden layers, the output spatial dimensions are specified at the bottom with the output spectral dimension on the top. 
		The dimensions of the input and output layer are also denoted.}
		\label{Fig:CNN1_25}
	\end{center}
\end{figure}

To implement networks, we use \texttt{TensorFlow}~\cite{TensorFlow}, an open-source machine learning platform
that contains \texttt{Keras}~\cite{Keras}, a deep learning application programming interface.
Since both the inputs and outputs are multi-channel images, it is natural to utilize only 2-D convolutional layers,
denoted by \texttt{Conv2D} in \texttt{Keras},
without applying flattening, which converts 2-D images or 3-D cubes into 1-D vectors as often done in CNN image classifications.
The network architecture is presented in Figure~\ref{Fig:CNN1_25}, where the output dimension for each layer are indicated.
The leftmost layer represents the input layer of dimension (150,~150,~9).
The input layer is followed by a sequence of multiple \texttt{Conv2D} layers without padding, each containing 25 kernels with varying kernel sizes.
As the input is passed through the network, the dimensionality is gradually decreasing toward (100,~100,~25), the dimension of the output layer.

For all \texttt{Conv2D}s, the convolution kernel~(filter) moves one pixel rightwards or downwards over a 3-D image between two successive applications of the kernel.
The kernel size for each \texttt{Conv2D} layer can be inferred from the difference between its output dimensions and that of the previous layer.
The first \texttt{Conv2D} layer, for instance, has kernel size (6,~6)~(height, width) that decreases the input of (150,~150,~9) into (145,~145,~25) in the absence of padding.
The kernel sizes are gradually reduced throughout the network. Overall, the CNN consists of 174,100 trainable parameters in total.
The motivation of such a tunnel-like architecture is to create a small network with relatively few parameters that features a short training time and fast predictions, namely a fast forward pass/propagation.
The network is referred to as CNN1, which reconstructs cubes~(network output) from CTIS images~(input). 

\begin{figure}[htp]
	\begin{center}
		\includegraphics[width=1\textwidth]{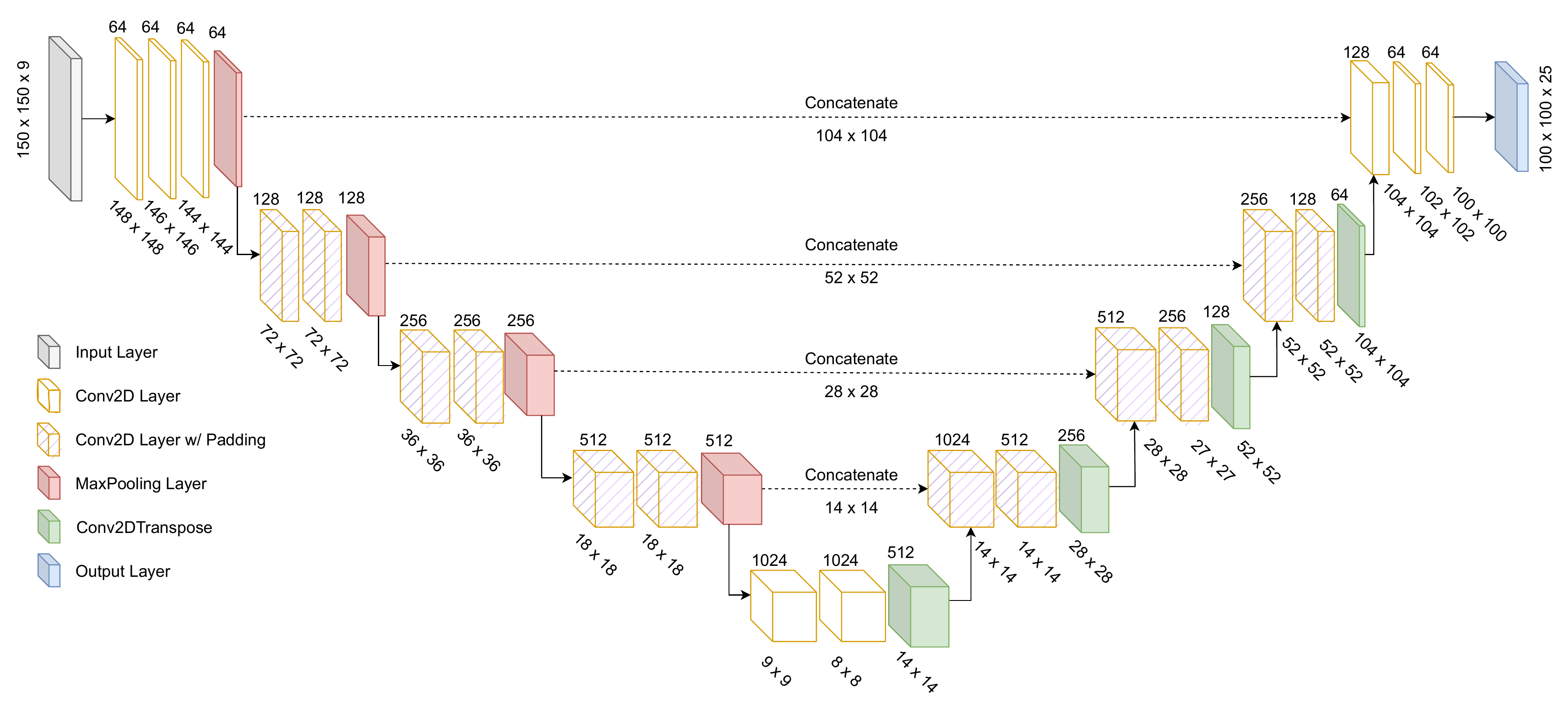}
		\caption{
			Illustration of the proposed UNet for reconstructing a $100\times100$ image with 25 spectral channels. The network has a total of 22,226,329 trainable parameters and consists of an input and output layer as well as layers of \texttt{Conv2D}, \texttt{Conv2D} with padding, \texttt{MaxPooling} and \texttt{Conv2DTranpose}.
			The output spatial dimensions of the layers are specified at the bottom with the output spectral dimension on the top.
			The skip-concatenation connections are indicated by the dotted arrows.
		}
		\label{Fig:UNET_25}
	\end{center}
\end{figure}

Lastly, we also test a U-Net~\cite{10.1007/978-3-319-24574-4_28}, which has been extensively used for image segmentation.
The U-Net architecture is adapted to match the dimensions of the input and output in question (Figure~\ref{Fig:UNET_25}) and consists of 22,226,329 trainable parameters, more than 100 times larger than the CNN1.

\subsection{Hybrid models}

From our previous work~\cite{huang2022application}, it has been demonstrated that neural networks can 
efficiently and accurately reconstruct cubes with smaller errors than the EM method, provided that the networks have been exposed to images of similar objects or geometries in the training phase. In addition, unlike the EM algorithm, the networks are not vulnerable to noise\footnote{We have found that for typical noise levels of our own CTIS camera, the network performance is not affected by the presence of noise. Furthermore, networks, that are exposed to noisy data, generalize slightly better to unseen data cubes, compared to those that see noiseless data only.}
and complex spatial variation in the images, giving rise to consistent results.

To improve the network's generalizability and overcome EM's weakness against noise and complex geometry, it is natural to combine the two methods sequentially. That is,  one first uses a network to reconstruct a cube from a CTIS image, which is passed to the iterative EM algorithm as an initial guess to further refine the network reconstruction.
Section~\ref{sec:results} details how the hybrid models produce better results than the EM algorithm and networks alone for both seen (cubes that are part of training and test sets) and unseen (neither part of training nor test set) data cubes.

It should be pointed out that Ref.~\cite{2017} has proposed to solve ill-posed inverse problems using iterative deep neural networks,
	where a known, traditional algorithm, which tackles inverse problems, is applied before neural networks.
	In the context of hyperspectal cube reconstruction, it corresponds to the reverse sequence: EM $\to$ network.
Because the network can handle the noise better than the EM,
the sequence we suggest, network $\to$ EM, will in principle provide a more consistent and stable reconstruction
with better performance than the reverse one as demonstrated in Section~\ref{sec:25_results}.
Moreover, for the sequence network $\to$ EM, one can experiment with different numbers of EM iterations to
attain an optimal balance between the performance and execution time.
By contrast, one must retrain the network in the EM $\to$ network framework once the number of iterations changes, 
as the input of the network is the outcome of the EM algorithm which depends on the number of iterations.
In other words, the network $\to$ EM has more flexibility in the implementation of real-world data.

For the reverse case, the EM outputs hyperspectral cubes which are fed into the CNN. Therefore, both the input and output of the CNN are cubes and thus have the dimensions $100\times100\times25$.
We use the same number of kernels and the same kernel sizes  as in CNN1 for a fair comparison.
To match the input and output dimensions, we use padding in each layer, to maintain the dimensionality throughout the network. This network consists of 188,500 trainable parameters, slightly more than the CNN1 due to padding, and is referred to as CNN2.

All in all, we have three hybrid models,  CNN $\to$ EM, EM $\to$ CNN and U-Net $\to$ EM which are denoted as
CNN1-EM, EM-CNN2 and UNet-EM, respectively.

\subsection{Network training \label{sec:data_CNN}}
For the network training procedure for all three networks, we choose the Adam optimizer, Mean-Squared-Error (MSE) as the loss function and (Keras)~\texttt{EarlyStopping Callbacks}, which ceases the training when the MSE of the validation set stops improving, to prevent overtraining. Moreover, we set the batch size to be 32 and 500 epochs are used, which are divided into 10, 10 and 480:
A learning rate of $4\times 10^{-5}$  is assumed for the first 10 epochs
but is reduced by a factor of 2 for the second 10 epochs and a factor of 4 for the following 480 epochs.
In addition, during the 480 epochs, the learning rate decays exponentially -- it is reduced by a factor of 0.9 for every 50,000 steps.
Moreover, 20 iterations are carried out for the  standalone EM algorithm but only 10 iterations for the EM step in the hybrid models.

To quantify the performance of the different methods, we utilize the MSE and peak signal-to-noise ratio (PSNR) in decibels as error metrics:

\begin{align}
	\text{MSE} = \frac{1}{N} \sum_{i=1}^{N}\left(Y_i-\hat{Y}_i\right)^2, \qquad \text{PSNR} = 10 \log_{10}\left(\frac{255^2}{\text{MSE}}\right)
\end{align}
where $N$ is the data sample size, $i$ is the sample index, $\hat{Y}_i$ is the reconstructed cube, and $Y_i$ is the ground truth. The maximum pixel value of the data cubes is 255 since the image format is 8-bit.

\subsection*{Noise influence}
Before presenting our results, we should point out that the noise term $\boldsymbol{n}$ in Eq.~\eqref{eq:gHf},
parametrized by a zero-mean Gaussian distribution, can affect the EM reconstruction performance~\cite{snyder_noise_1987}.
Zeng et al.~\cite{zeng_unmatched_2000} investigated the ill-conditioned image reconstruction problem in the presence of noise and showed that the EM algorithm 
at the beginning demonstrates a short convergent trend but then diverges from the desired solution.


Empirically, for zero-mean Gaussian noise with a standard deviation of 0.5 at a maximal pixel value of 255, i.e. a noise level of $\approx 0.2 \, \%$,
the application of the EM after neural networks sometimes increases MSE instead of refining the network predictions which is similar to the behavior observed by Zeng et al.~\cite{zeng_unmatched_2000}. This is especially the case for network predictions that are very close to the ground truth, where the effect of a small mismatch is accumulated during the EM iterations,
which involve matrix multiplications between tensors of large dimensions, and perturbs the reconstructed cubes away from the true ones. The effects of noise on our models are briefly investigated in Appendix \ref{app:Noise}.
%

%

For simplicity, in this work we do not consider the noise term when generating the data,
assuming the noise is either small enough or can be included into cubes $\boldsymbol{f} $.

\section{Comparison of model performance and extension toward hyperspectral regimes}\label{sec:results}

In this section, we first present our results by comparing the different reconstruction approaches.
Second, we investigate whether the hybrid approach can be applied in a more challenging hyperspectral scenario with 100 spectral channels.

\subsection{Results of 25 spectral channels}\label{sec:25_results}
The results of applying the training procedure explained above and the EM algorithm are summarized in Table~\ref{tab:25Results}.
Based on the results we make the following observations and comments.

\begin{table}[htp]
	\centering
	\begin{tabular}{P{2cm}<{\raggedright} S[table-column-width=5cm] S[table-column-width=5cm] S[table-column-width= 1.6cm]}
		\toprule
		& {Seen Cubes~(19,665 samples)} & {Unseen Cubes~(805 samples)} & {Time~(ms)}\\
		\midrule
		EM & 121.50~\!\!\!\!(27.3) & 153.04~\!\!\!\!(26.3) & 48.05 \\ 
		CNN1 & 10.88~\!(37.8) & 22.37~\!(34.6) & 0.90 \\
		CNN1-EM & 7.91~\!(39.2) & 16.67~\!(35.9) & 26.31 \\ 	
		EM-CNN2 & 8.45~\!(38.9) & 18.63~\!(35.4) & 26.06 \\ 
		U-Net & 0.91~\!(48.6) & 10.34~\!(38.0) & 1.69 \\
		UNet-EM & 0.78~\!(49.2) & 8.83~\!(38.7) & 27.06 \\
		\bottomrule
	\end{tabular}
		\caption{MSE~(PSNR) for the models under consideration as well as the EM algorithm, where the test set of the seen (19,665) samples and unseen (805) samples are used to evaluate the model performance. Table also shows the average computation times for the respective models}	
	\label{tab:25Results}
\end{table}

\begin{itemize}
\item The hybrid models perform better than both the standalone EM and networks:
$\sim 26~\%$ improvement on CNN1-EM compared to CNN1, and $\sim 14~\%$ improvement on UNet-EM compared to U-Net. The improvement occurs for both seen and unseen cubes,
while the EM alone is by far worse than the network models.   
It illustrates the strength of the sequential combination of the network and EM -- the networks provide a good initial condition for the EM to further improve.

\item The EM results for the seen and unseen cubes, MSE 122.50 versus 153.04, indicate the performance inconsistency associated with the spatial variation in the data.
In fact, the fluctuation in MSE is even more pronounced when comparing samples from different individual pushbroom cubes. 

\item The model CNN1-EM outperforms EM-CNN2. That corroborates our previous argument that the EM is more vulnerable
to noise and complex geometry and might yield inconsistent results as inputs to the network.
In this case, the CNN2 must cope with different degrees of variation from the EM output
and overall  performs worse than the CNN1-EM.

\item By contrast, in the CNN1-EM model,  the vulnerability of EM has been mitigated by the network which
decreases the noise and supplies a good starting point for the EM.
Moreover, one can freely experiment with different numbers of EM iterations in the CNN1-EM to attain an optimal balance between the accuracy and reconstruction time,
whereas the CNN2 must be retrained whenever the number of iterations is changed in the EM step.

\item The U-Net as a much bigger network~(more than 100 times larger than CNN1) has a more significant MSE increase, a factor of 11, from the seen to unseen cubes
while the MSE only doubles for CNN1. The U-Net has, nonetheless, a smaller MSE than CNN1 for the unseen cubes.
It illustrates first that it is very important to include different objects with various geometries into the training set for  the network, especially large ones, 
to maintain consistent performance over all different objects.
Second, a smaller network is more robust against new data and performs more consistently than a large network. 

\end{itemize}

\begin{figure}[htp]
	\begin{center}
			\includegraphics[width=1\textwidth]{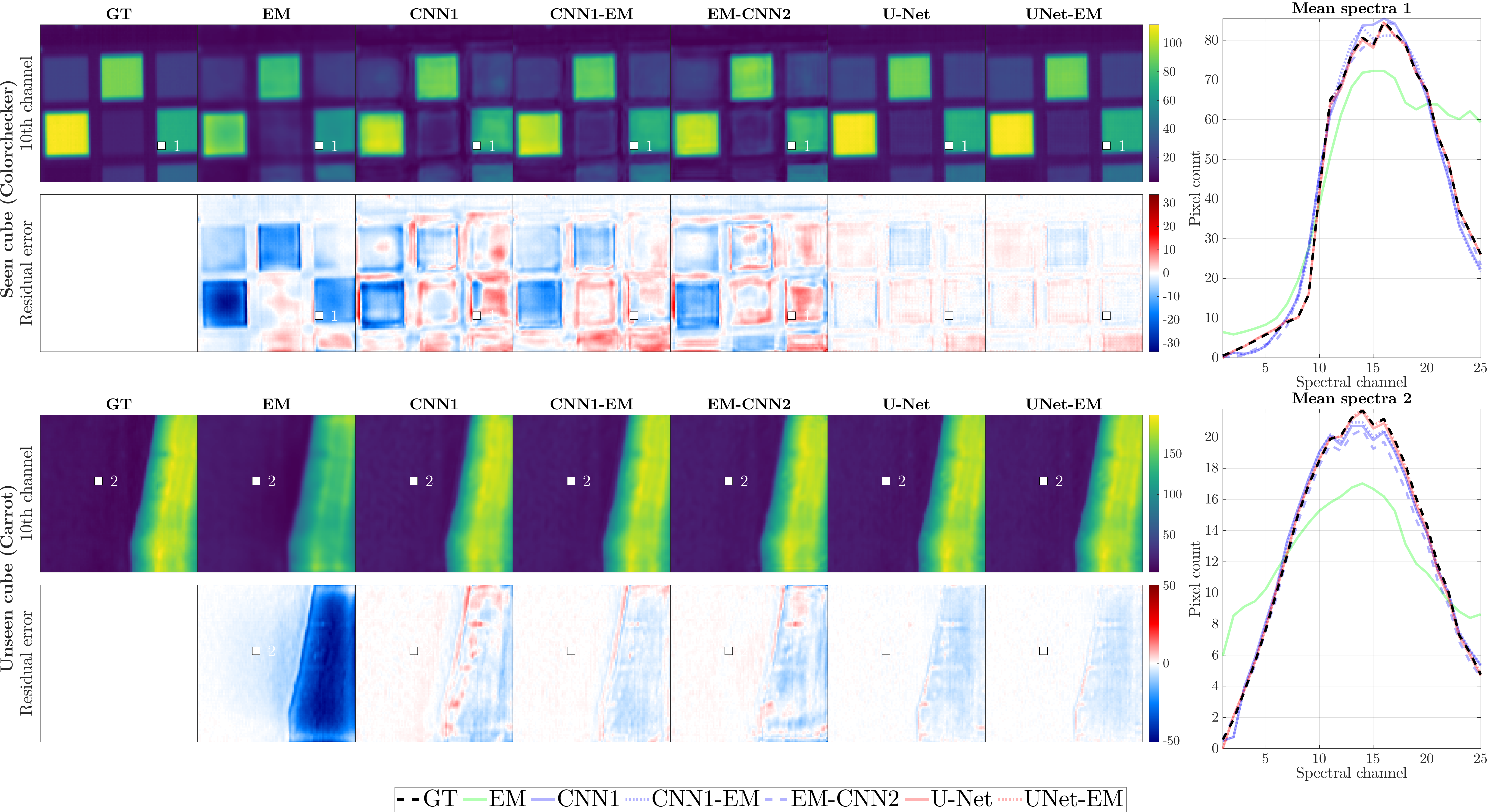}
			\caption{Comparison of 25-channel reconstructions of the seen (Colorchecker) and unseen (carrot) cube for the Ground Truth~(denoted by GT), EM~(20 iterations), CNN1,
			CNN1-EM~(10 iterations), EM-CNN2, U-Net and UNet-EM. The 10th spectral channel and the corresponding residual error (reconstructed - GT) are shown in the images, while the mean $5\times 5$~pixels (represented by the white squares in the images) spectra are also plotted.}
			\label{Fig:100x100x25_results}
		\end{center}
\end{figure}

In Figure~\ref{Fig:100x100x25_results}, we show the 10th channel of the Ground Truth~(referred to as GT) and reconstructed cubes for the seen~(top) and unseen~(bottom) cubes
as well as the residual error.
The UNet-EM~(EM alone) has the best~(worst) accuracy as seen from residual errors.
The spectral information, characterized by the mean pixel value of a representative
$5\times5$ area denoted by the white square in the images, is also shown. That is, the mean pixel value as a function of spectral channels.
The UNet-EM~(pink dotted line) follows the GT best, while EM is unable to reproduce the spectral shape in detail.
The advantage of applying EM after the networks  are more visible for the unseen cubes.
For example, around the middle channels it moves the spectra closer to the true ones.
Additional figures for different cubes are shown in Figures~\ref{Fig:100x100x25_results_02-03}-\ref{Fig:100x100x25_results_04} in Appendix \ref{app:25_channel}.
Finally, to easily visualize how closely the reconstructed cubes resemble the true ones, we show the RGB images of some representative reconstructed cubes in Supplemental Material in Section~S7.

\subsection{Results of 100 spectral channels}

We are now in a position to tackle a more challenging scenario of 100 spectral channels with the hybrid approach.
Only a U-Net network (modified to match dimensions accordingly) of 22.2 million parameters is considered as it has the best performance among the network models for 25 channels.
We follow closely the data-creation and training procedure used in the 25-channel case with one major difference --
100 epochs are assumed instead of 500 for the network training to reduce the training time
as it takes a much longer time for each epoch with 100 spectral channels.
Finally, we apply the EM algorithm with 10 iterations to further refine the network predictions.

The performance of the standalone and hybrid models are summarized in Table~\ref{tab:100_comp}.

\begin{table}[htp]
	\centering
	\begin{tabular}{P{2.2cm}<{\raggedright} S[table-column-width=5.9cm] S[table-column-width=5.9cm] }
		\toprule
		& {Seen Cubes~(19,665 samples)} & {Unseen Cubes~(805 samples)} \\
		\midrule
		EM & 27.62~(33.7) & 45.99~(31.5) \\
		U-Net & 3.51~(42.7) & 19.80~(35.2) \\
		UNet-EM & 2.83~(43.6) & 11.80~(37.4) \\
		\bottomrule
	\end{tabular}
\caption{MSE~(PSNR) for the EM algorithm, U-Net, and hybrid network. Similar to Table~\ref{tab:25Results}, the test set of the seen (19,665) samples and unseen (805) samples are used to evaluate the model performance.}
	\label{tab:100_comp}
\end{table}
For the seen cubes, the hybrid model is much better than the EM algorithm
by a factor of 9.7 in MSE while the hybrid model is better than the U-Net by $19~\%$.
For unseen cubes, although the performance of the U-Net significantly decreases,
it still outperforms the EM.
The EM step in the UNet-EM improves the U-Net predictions by $40~\%$, which is significantly higher than the 25-channel case~(only  $15~\%$).
It implies the EM step is more pivotal in the hyperspectral regime, especially for the completely unseen data.
 
%
\begin{figure}[h]
	\centering
	\includegraphics[width=\textwidth]{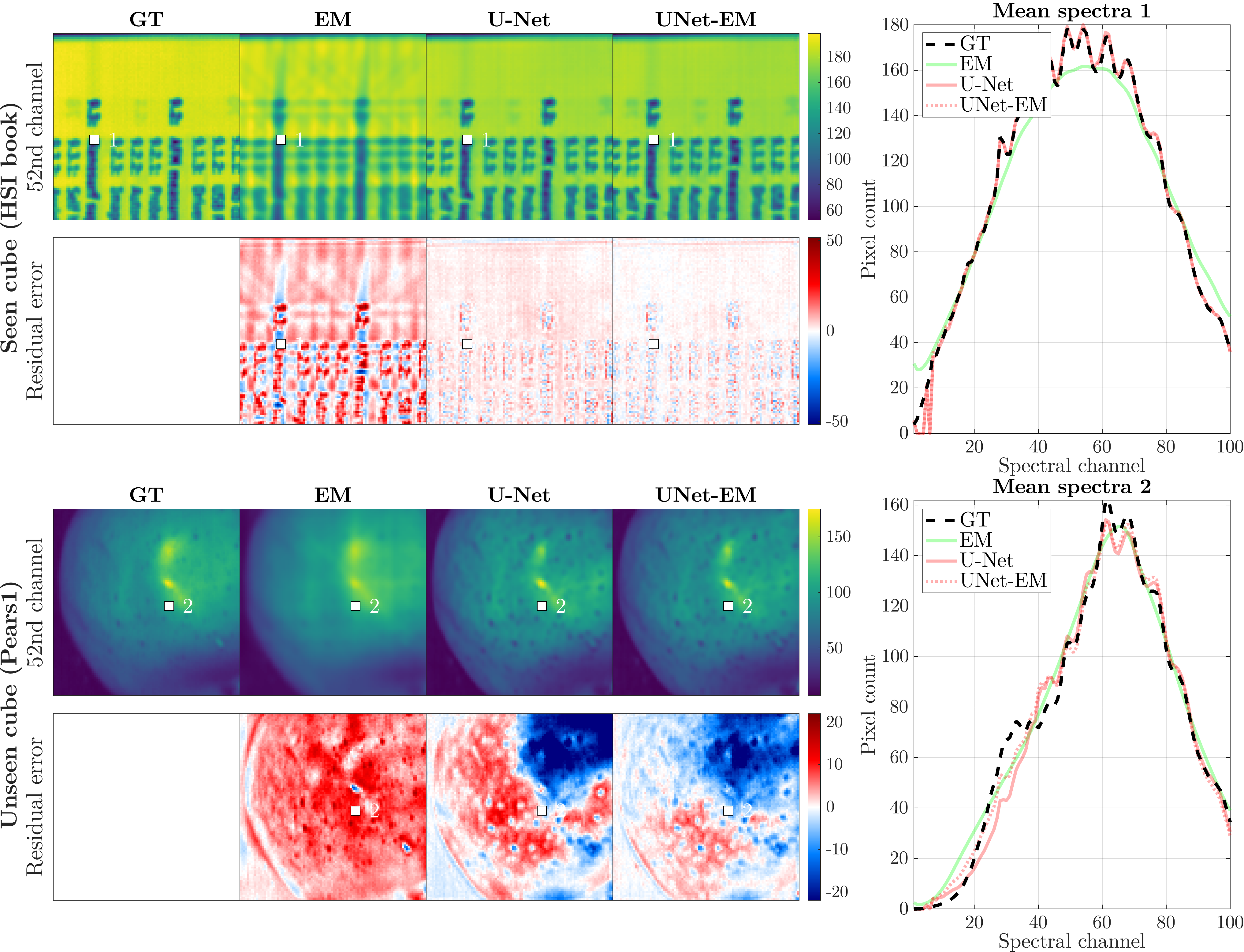}
	\caption{Comparison of 100-channel reconstructions of the seen (HSI book) and unseen (Pears1) cube for GT, EM (20 iterations), U-Net and UNet-EM. The 52nd spectral channel and the error (reconstruction-GT) for the respective reconstructions are shown in the images, while the mean $5\times 5$~pixels (indicated by the white squares) spectra are shown in the plots.}
	\label{Fig:100x100x100_results}
\end{figure}

Similar to Figure~\ref{Fig:100x100x25_results}, we show the 52nd channel of true and reconstructed cubes as well as
the spectral information in Figure~\ref{Fig:100x100x100_results} for a seen and unseen cube.
The EM alone can only reconstruct the overall spectral shape but again fails to capture small variations and details as seen by comparing the black~(ground truth) and green~(EM) lines
in the spectral plots. From the 52nd channel it is also evident that spatial information is lost for EM, where the text from the seen cube is no longer readable~(top row of the left panel in Figure~\ref{Fig:100x100x100_results}) and the finer spatial details of the unseen cube are missing. This smearing or smoothing of the high-frequency spatial and spectral components is characteristic of the EM algorithm. Moreover, the EM in the UNet-EM visibly reduces the level of residual errors with respect to the U-Net predictions. Additional figures for different cubes are shown in Figures~\ref{Fig:seen_vs_unseen_02-03}-\ref{Fig:seen_vs_unseen_04} in Appendix \ref{app:100_channel}.
Similar to the 25 channels case, to visualize how closely the reconstructed cubes mimic the true ones,
we show RGB visualizations of some representative reconstructed cubes in Section~S8 in Supplemental Material.

To summarize, we have demonstrated that the hybrid model can make very good predictions on a variety of cubes for 100-channels -- a hyperspectral regime --
and can decently generalize to totally unseen samples.
It outmatches both the standalone U-Net and the EM algorithm.
The improvement by including the EM~(compared to the U-Net alone) is more pronounced than in the 25-channel case:
$19 - 40~\%$ for 100 channels versus $\sim 14~\%$ for 25 channels.
The hybrid model has proved promising with broad applications in the reconstruction of real-world hyperspectral imaging.

\section{Conclusions}\label{sec:conclusion}

The CTIS, a snapshot hyperspectral imaging system, is a compact and efficient way of providing hyperspectral information. A 3-D hyperspectral cube can be reconstructed from a CTIS image that entails wider applications than the 2-D image itself.
It has been shown~\cite{huang2022application} that CNNs can be employed for fast, reliable cube reconstruction, provided that the CNNs have been exposed to
objects of similar geometry during the training.
On the other hand, the iterative reconstruction algorithms, e.g. EM, need no {\it training} and can be applied to different CTIS images.

In this work, we propose a very simple but novel way of cube reconstruction -- a hybrid model.
The model first utilizes a network to reconstruct a hyperspectral cube from a CTIS image, and the reconstructed cube is fed into the EM algorithm as an initial value of the cube, which is then recursively updated. 
We have trained and tested our hybrid models based on real-world hyperspectral cubes from a pushbroom camera and CTIS images,
generated by applying the cubes to a realistic CTIS simulator. The simulator~(See Supplementary information) emulates a real CTIS system based on experimental measurements of the point-spread-function, illumination and diffraction sensitivity as a function of the wavelength.
We studied scenarios of 25 and 100 spectral channels. For both of the scenarios, the data consist of training~(91998 samples) validation~(19665), and test~(19665) sets, cropped from 171 different Pushbroom cubes.
The performance of models is evaluated  based on the test set as well as an extra 805 samples, created from 7 unseen pushbroom cubes.
For comparison, we investigated different methods of reconstruction.

For 25 spectral channels, we consider the standalone EM, CNN, U-Net and hybrid models of CNN1-EM~(EM is applied after CNN),
EM-CNN2~(CNN is applied after EM), UNet-EM,
where the U-Net, CNN1  and CNN2  have 22.23, 0.17 and 0.19 million parameters, respectively.
The performance of the models is summarized in Table~\ref{tab:25Results}.
First, it has been found that UNet-EM is the best model whereas EM alone has the worst performance. That shows the advantages of neural networks
over the traditional reconstruction algorithm, as demonstrated in our previous work~\cite{huang2022application}, provided that the networks have seen a variety of different objects.
Second, all hybrid models perform better than the corresponding networks alone with the improvement ranging from $14~\%$ to $27~\%$.
In addition, the CNN1-EM outperforms the reverse order EM-CNN2. It highlights the synergy between the networks and EM as follows:
The EM can help networks to cope with unseen data as it can be applied to images of any objects.
On the other hand, the network is less prone to noise and provides a good and stable initial guess for the EM to further improve the results.
The reverse order EM-CNN2 will, by contrast, be subject to the noise and inconsistent results from the EM.
Finally, the U-Net as a much larger network experiences a more noticeable performance loss~(roughly a factor of 10)
from the seen to unseen cubes as opposed to the much smaller CNN1~(a factor of 2), although the U-Net still reconstructs cubes better.
It indicates that a smaller network is more robust against new types of cubes.
Additionally, the smaller network will also have a shorter forward pass time, i.e., faster predictions.
Therefore, one should find the optimal configuration by considering the reconstruction time, performance and robustness when advancing to the real-time reconstruction.

We have also demonstrated that the hybrid approach works well in the hyperspectral regime by exploring a 100-channel case. The results are summarized in
Table~\ref{tab:100_comp}. The inclusion of the EM significantly improves the U-Net result, especially for the unseen data -- the MSE is reduced by nearly a factor of 2.

To summarize, we have presented a very simple but novel hybrid model of hyperspectral cube reconstruction by applying the traditional EM algorithm after neural networks.
These two methods are complementary -- the network, which is less susceptible to noise and complex spatial variation in the images, provides a refined initial condition for the EM while the EM further improves the network's results and assists it to deal with very different kinds of objects.

\section*{Acknowledgements}
MTF and WCH acknowledge partial funding from The Villum Foundation and a CenSec grant funded by The Danish Ministry of Higher Education and Science (CenSec). MSP acknowledge partial funding from the Innovation Fund Denmark (IFD) under File No. 1044-00053B.
We acknowledge partial support from Food $\&$ Bio Cluster Denmark.
This work was performed using the \href{https://escience.sdu.dk/index.php/ucloud/}{UCloud} computing and storage resources, managed and supported by eScience center at University of Southern Denmark.

\newpage
\begin{appendix}
\csname efloat@restorefloats\endcsname

\section{Additional comparisons of the different reconstructions - 25 channels}\label{app:25_channel}
\begin{figure}[h!]
	\begin{center}
		\includegraphics[width=\textwidth]{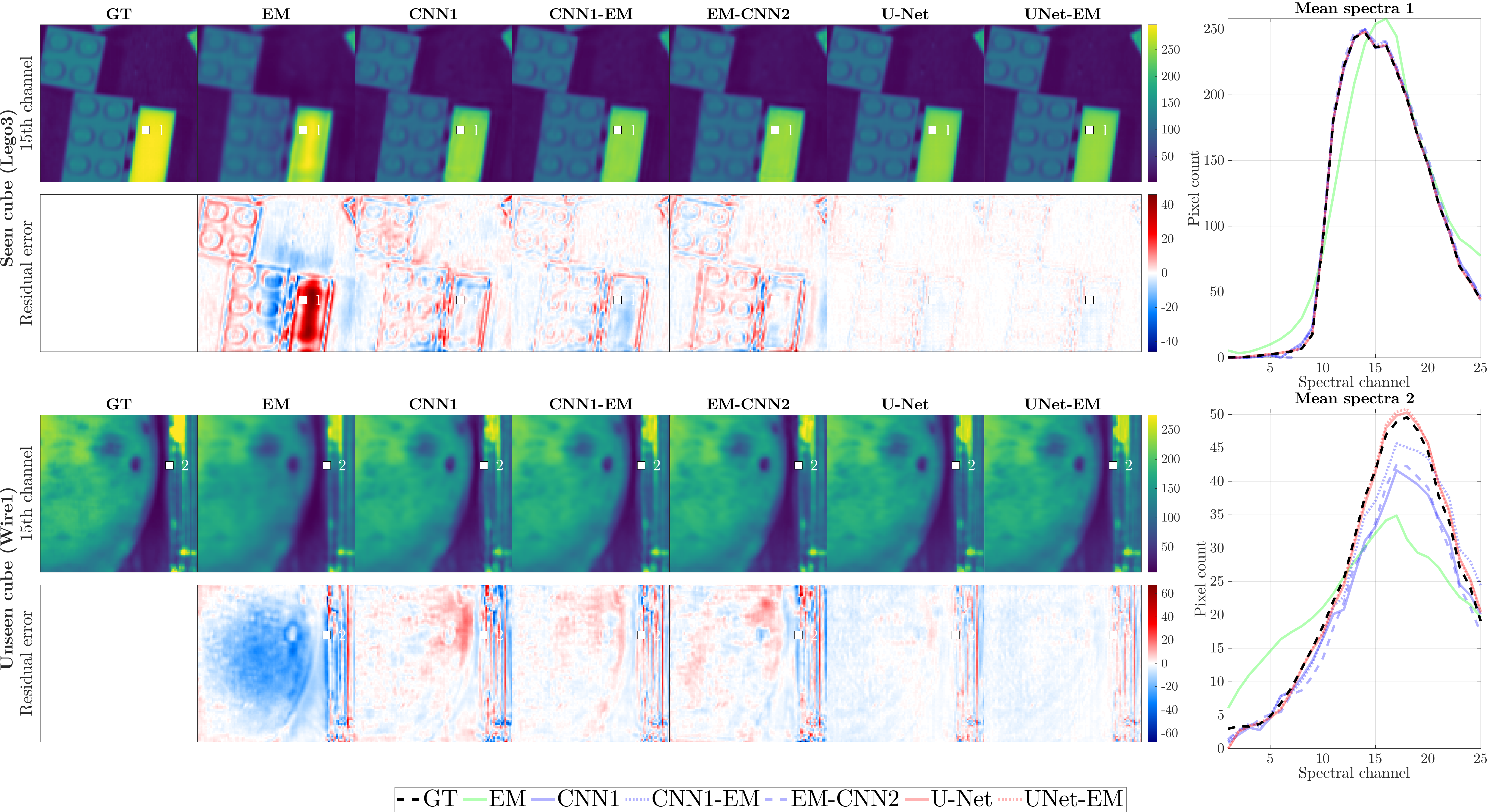}
		\caption{Comparison of 25-channel reconstructions of the seen (Lego3) and unseen (Wire 1) cubes for GT, EM (20 iterations), CNN1, CNN1-EM, EM-CNN2, U-Net and UNet-EM. The 15th and 17th spectral channels and the error (reconstruction-GT) for the respective reconstructions are shown in the images, while the mean $5\times 5$~pixels (indicated by the white squares) spectra are shown in the plots.}
		\label{Fig:100x100x25_results_02-03}
	\end{center}
\end{figure}
\begin{figure}[h!]
	\begin{center}
		\includegraphics[width=\textwidth]{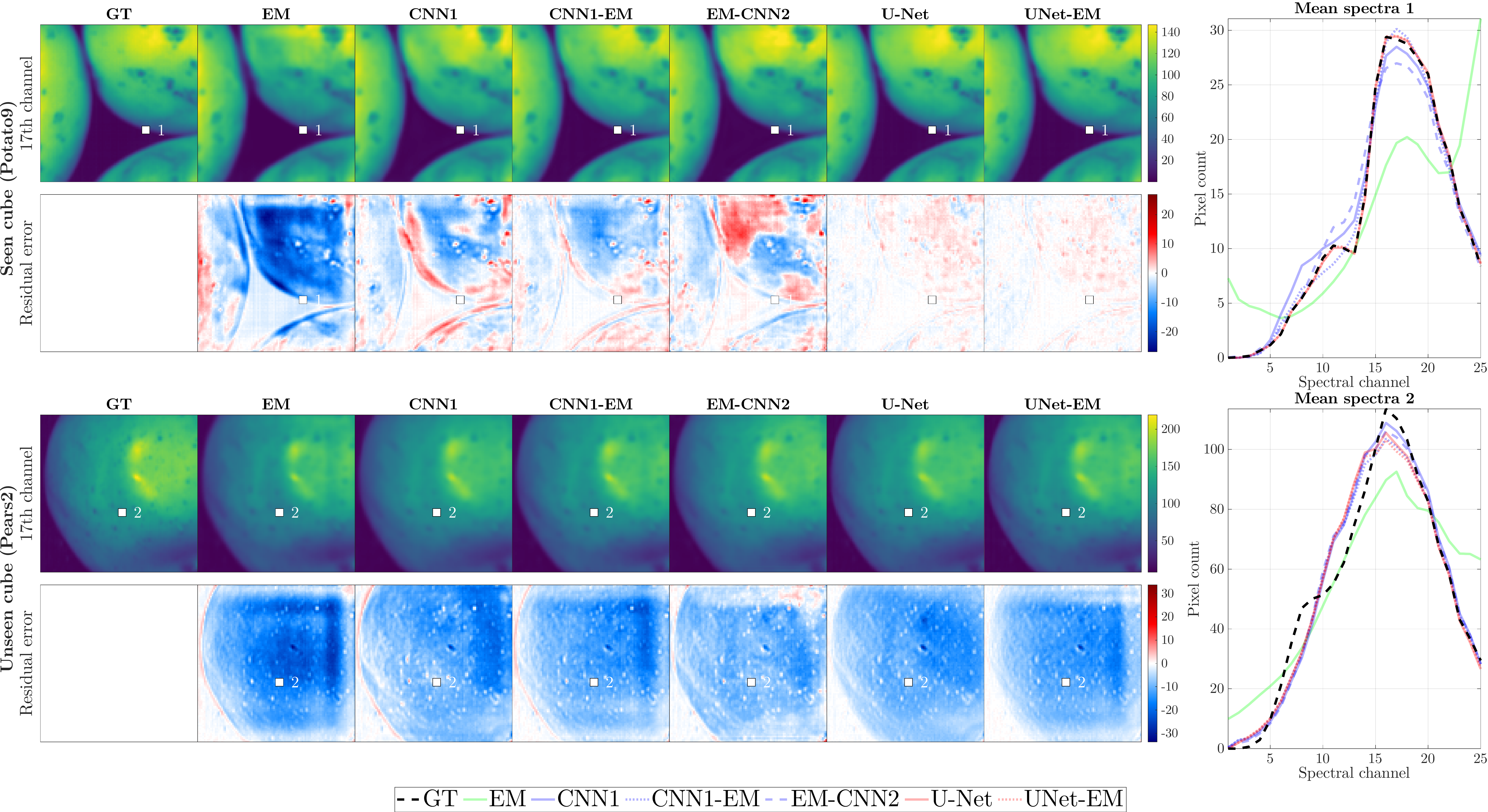}
		\includegraphics[width=\textwidth]{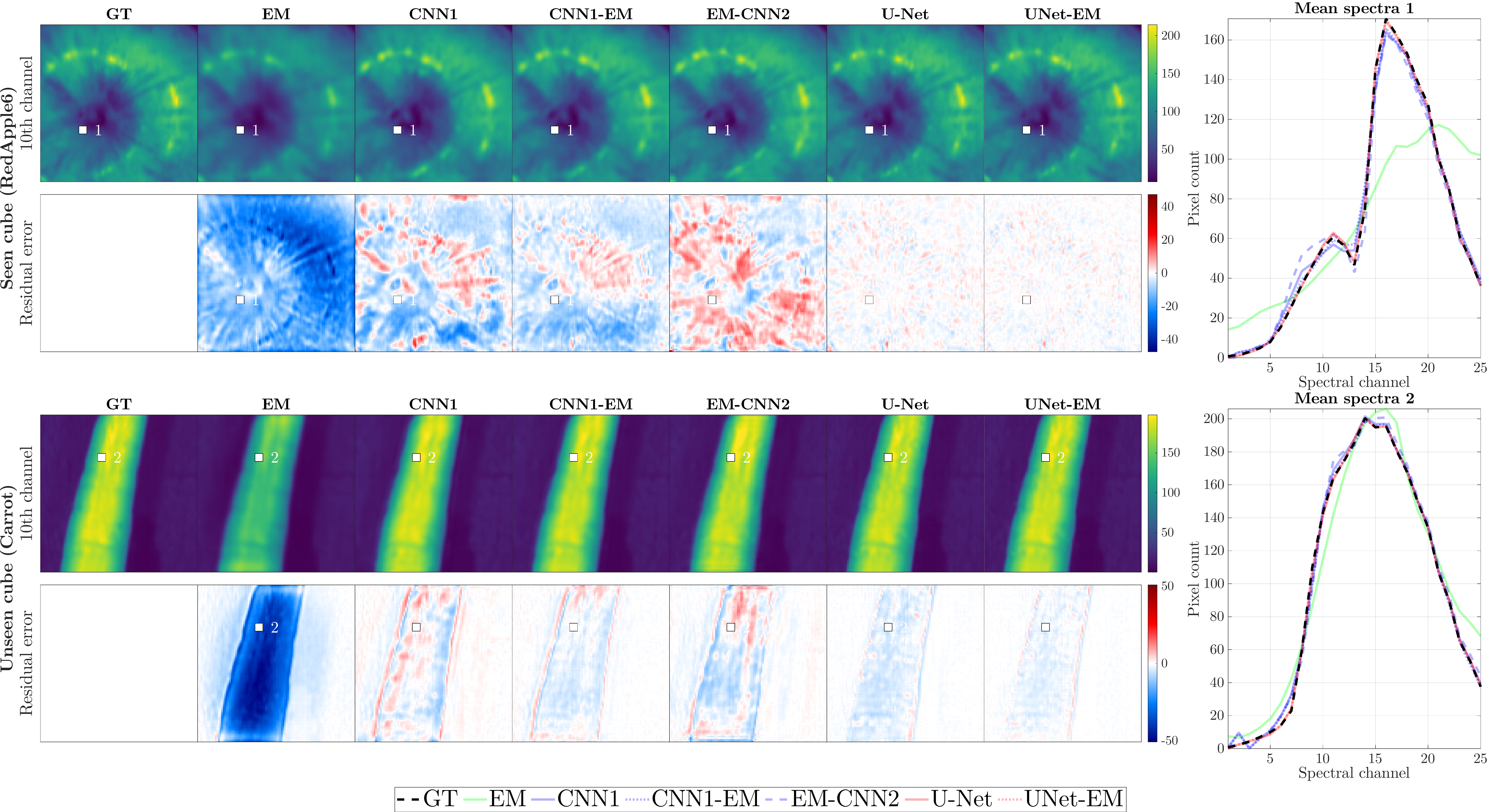}
		\caption{Comparison of 25-channel reconstructions of the seen (Potato9 and RedApple6) and unseen (Pears2 and Carrot) cube for GT, EM (20 iterations), CNN1, CNN1-EM, EM-CNN2, U-Net and UNet-EM. The 10th spectral channel and the error (reconstruction-GT) for the respective reconstructions are shown in the images, while the mean $5\times 5$~pixels (indicated by the white squares) spectra are shown in the plots.}
		\label{Fig:100x100x25_results_04}
	\end{center}
\end{figure}

\clearpage
\section{Additional comparisons of the different reconstructions - 100 channels}\label{app:100_channel}
\csname efloat@restorefloats\endcsname

\begin{figure}[h!]
	\centering
	\includegraphics[width=\textwidth]{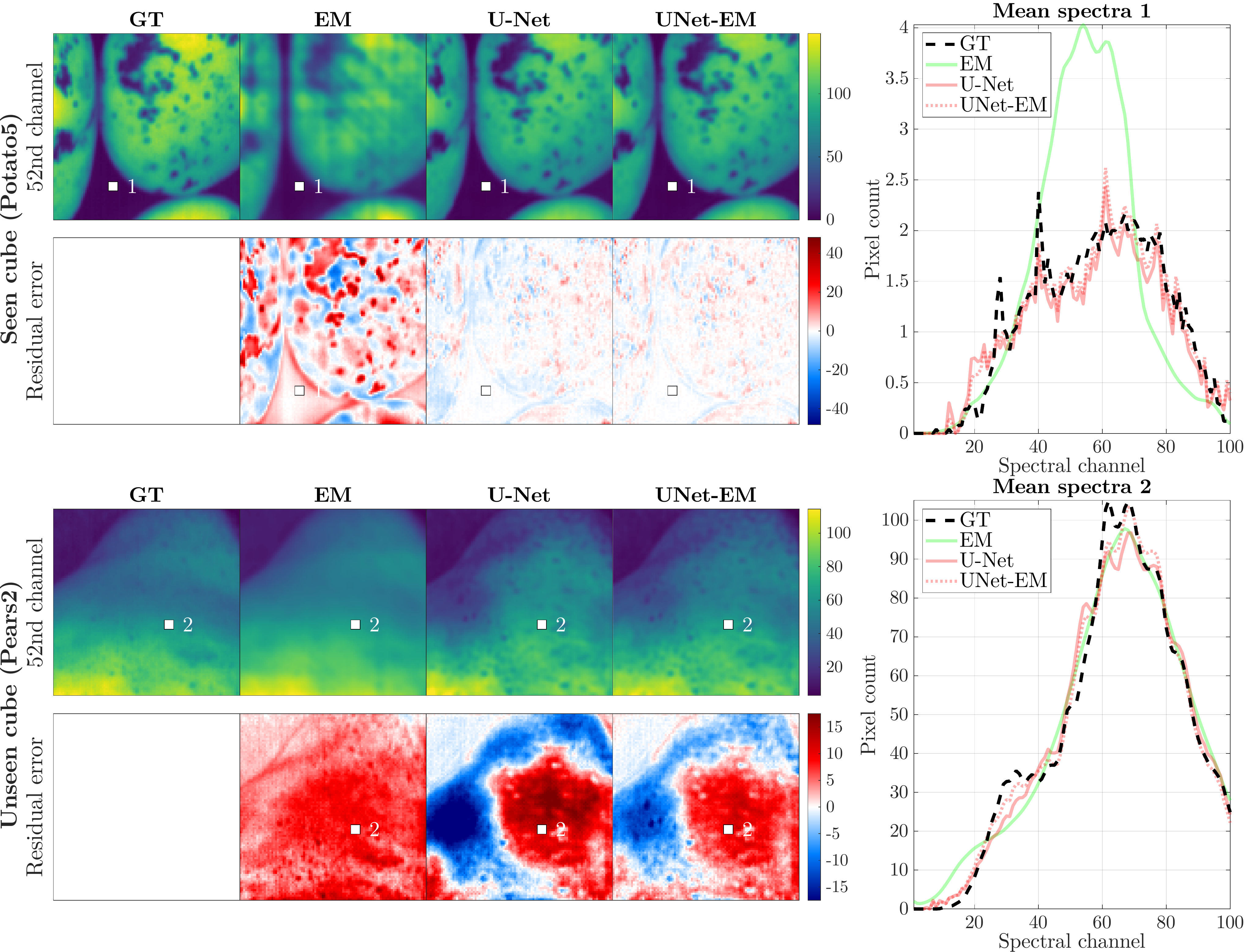}
	\caption{Comparison of 100-channel reconstructions of the seen (Potato5) and unseen (Pears2) cube for GT, EM (20 iterations), U-Net and UNet-EM). The 52nd spectral channel and the error (reconstruction-GT) for the respective reconstructions are shown in the images, while the mean $5\times 5$~pixels (indicated by the white squares) spectra are shown in the plots.}
	\label{Fig:seen_vs_unseen_02-03}
\end{figure}

\begin{figure}[h!]
	\centering
	\includegraphics[width=\textwidth]{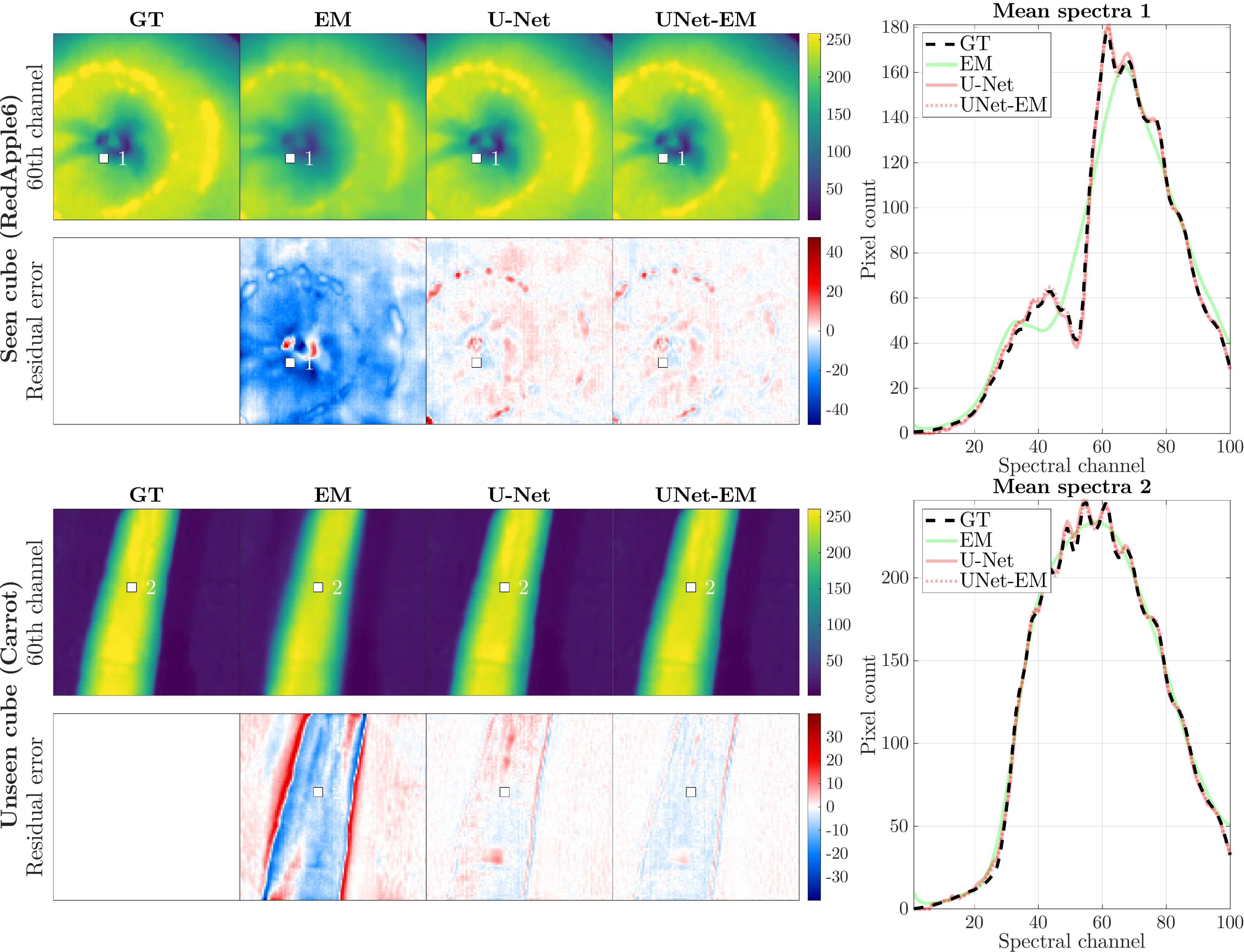}
	\caption{Comparison of 100-channel reconstructions of the seen (iPhone8) and unseen (Wire1) cube for GT, EM (20 iterations), U-Net and UNet-EM. The 30th spectral channel and the error (reconstruction-GT) for the respective reconstructions are shown in the images, while the mean $5\times 5$~pixels (indicated by the white squares) spectra are shown in the plots.}
	\label{Fig:seen_vs_unseen_03}
\end{figure}
\begin{figure}[!]
	\centering
	\includegraphics[width=\textwidth]{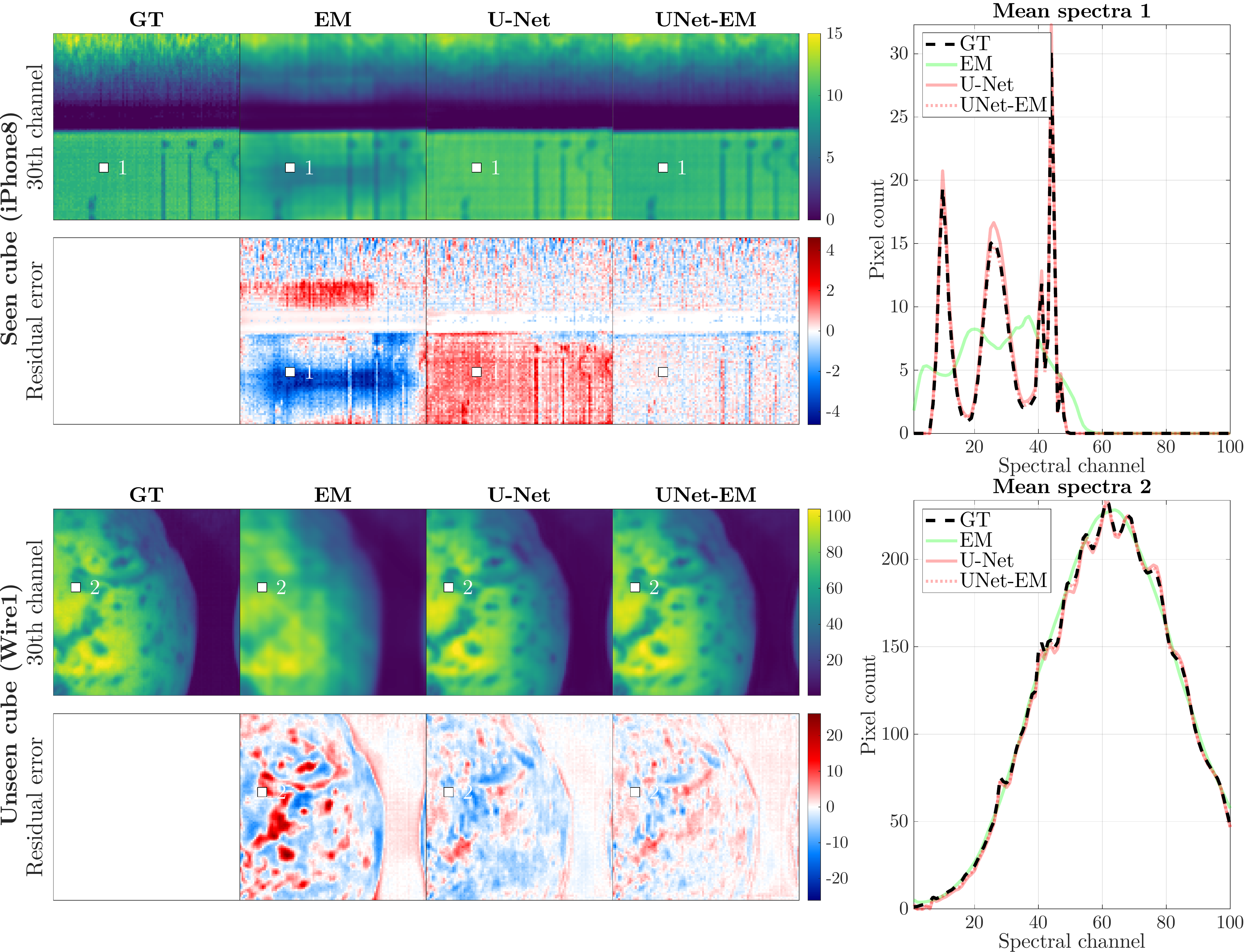}
	\caption{Comparison of 100-channel reconstructions of the seen (HSI book) and unseen (Pears1) cube for GT, EM (20 iterations), U-Net and UNet-EM. The 52nd spectral channel and the error (reconstruction-GT) for the respective reconstructions are shown in the images, while the mean $5\times 5$~pixels (indicated by the white squares) spectra are shown in the plots.}
	\label{Fig:seen_vs_unseen_04}
\end{figure}

\clearpage

\section{Noise investigations}\label{app:Noise}
The following is a brief investigation into the effects of noise on the proposed hybrid approach.
To quantify the effects of noise, white Gaussian noise is added to the training, validation and test data, which consists of both seen and unseen cubes, and we here focus only on the hybrid CNN1-EM model.
The applied noise is characterized by a zero-mean Gaussian distribution with a standard deviation of $\sigma = 0.5$~(approximately corresponds to the real noise level found on our CTIS camera).
All negative pixel values resulted from the addition of the Gaussian noise are replaced by zero.
Notice that the added noise is \it{not} incorporated in the $\Hmat$ matrix of the EM algorithm.

We have found that the performance of CNN1 is not affected by the presence of the noise. Instead, in some cases CNN1 performs better on unseen cubes when trained on the noisy data.
In other words, the CNN benefits from being exposed to noise and thus becomes more robust. This is already a known feature of the convolution neural networks~\cite{nazare2017deep,7552554}.

\begin{figure}[h!]
	\centering
	\includegraphics[width=\textwidth]{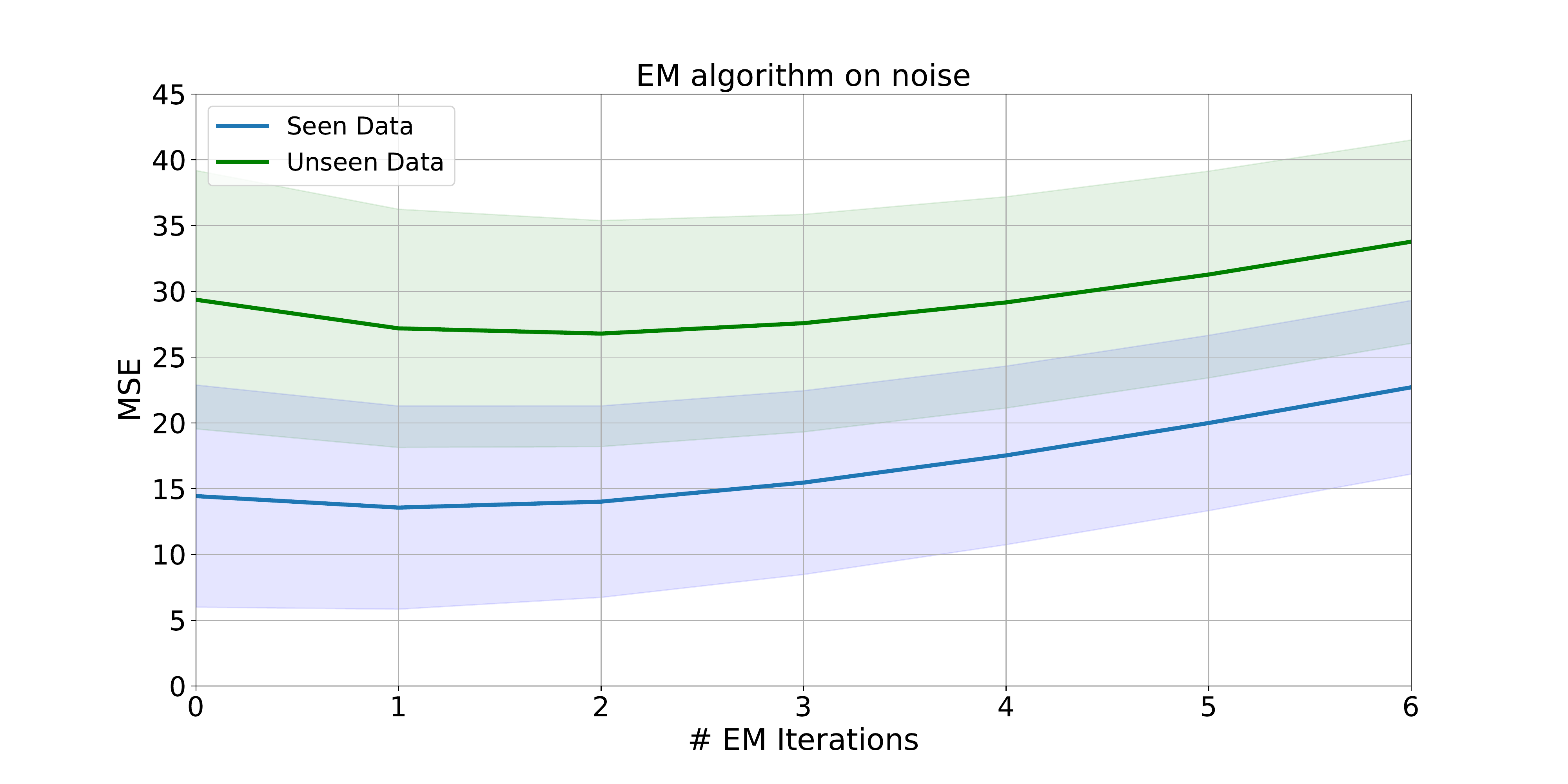}
	\caption{Visualization of the MSE of the EM algorithm as a function of iterations. The blue curve is evaluated on 4788 seen datasets, while the green curve is evaluated on 196 unseen datasets.}
	\label{Fig:Noise_Seen_vs_Unseen}
\end{figure}

On the other hand, the performance of CNN1-EM on the noisy data is shown in Fig.~\ref{Fig:Noise_Seen_vs_Unseen}, which
successfully reproduces the converging -- followed by the diverging -- behavior observed by Zeng et al.~\cite{zeng_unmatched_2000}. 
Furthermore, the quality of  initial guesses provided by CNN1 also has an impact on the performance of the second EM step.
With a \textit{decent} initial guess (close to the ground truth such as the blue line in Fig.~\ref{Fig:Noise_Seen_vs_Unseen} for seen cubes), the EM results will diverge quickly,
whereas with a \textit{mediocre} initial guess (such as the green line for unseen cubes) the EM results stay convergent for a longer time.

In fact, the existence of sizable noise implies a noticeable mismatch between the real CTIS image and the approximated, $\boldsymbol{\hat{g}} = \Hmat \boldsymbol{\hat{f}}$.
In this case, the $\Hmat$ matrix used in the EM algorithm does not correctly map the hyperspectral cubes to the CTIS images -- that
violates the assumption on which the EM algorithm (described in Eq.~\eqref{eq:EM_1}) is based, thus making the EM unable to attain the true cubes.

There are at least two ways to circumvent the noise issue.
First, one can properly model systematic errors of the CTIS camera and subtract them from the CTIS image such that the remaining random noise is small and under control. Alternatively, one could capture more than one CTIS image of the same object, and average out the systematic errors.
Second, the cube $\boldsymbol{f}$ in Eq.~\eqref{eq:gHf} can be redefined to include the random noise:
\[
\boldsymbol{g} = \boldsymbol{H}\boldsymbol{f} + \boldsymbol{n} \equiv  \boldsymbol{H}\boldsymbol{ \hat{f}} \; .
\]
In this case, one can obtain an estimated cube $\boldsymbol{ \hat{f}}$, which contains the noise, by the EM algorithm in a consistent way.
As long as the noise term is small enough, e.g. the standard deviation of the noise is less than one pixel count, the reconstructed cubes will be a good approximation to the real cube $\boldsymbol{f}$.

Conclusively, to utilize the proposed hybrid models, one has to very carefully model $\Hmat$, such that the level of noise is under control.


\end{appendix}

\clearpage

\bibliographystyle{unsrt}
\bibliography{ref_ref.bib}

\end{document}


\title{Supplemental material \\ The hybrid approach -- Convolutional Neural Networks and Expectation Maximization Algorithm -- for Tomographic Reconstruction of Hyperspectral Images}
	
	\date{}
	
	\author[,1]{\small{Mads Juul Ahleb{\ae}k}\footnote{ahle@sdu.dk}}
	\author[,2,3]{\small{Mads Svanborg Peters}\footnote{mape@newtec.dk}}
	\author[,1]{\small{Wei-Chih Huang}\footnote{huang@cp3.sdu.dk}}
	\author[,1]{\\ \small{Mads Toudal Frandsen}\footnote{frandsen@cp3.sdu.dk}}
	\author[,3]{\small{Ren\'e Lynge Eriksen}\footnote{rle@mci.sdu.dk}}
	\author[,2]{\small{Bjarke J{\o}rgensen}\footnote{bjarke@newtec.dk}}
	
	\affil[1]{\small{CP$^3$-Origins, Department of Physics, Chemistry and Pharmacy, University of Southern Denmark, Denmark}}
	\affil[2]{\small{Newtec Engineering A/S, 5230 Odense, Denmark}}
	\affil[3]{\small{Mads Clausen Institute, University of Southern Denmark, Denmark}}
	
	\maketitle
	\section{Introduction}
	This supplemental material accompanies the main paper and details the computed tomography imaging spectrometer (CTIS) simulator as well as the measurement of the parameters used in the simulator. Additionally, RGB visualizations of all the hyperspectral images captured with our VIS-NIR laboratory pushbroom system are shown. A selection of the reconstructed hyperspectral images for both the 25- and 100-channel cases are also compared through RGB visualizations. 
	\section{Computed tomography imaging spectrometer \\ simulator}\label{sec:simulator}
	The CTIS simulator introduced in our work is an updated version of the simulator used in our previous work \cite{huang2022application}, and is used to create input CTIS images for our network. This updated simulator includes zero mean white Gaussian noise, a point spread function (PSF) and spectral sensitivity corrections due to the illuminant and the diffractive optic element (DOE). Thus, the simulator requires the following inputs; \texttt{x}, \texttt{y}, \texttt{z}, \texttt{b1}, \texttt{b2}, \texttt{shift}, \texttt{allOrders}, \texttt{diff\_sens}, \texttt{illum}, \texttt{sigma\_psf} and \texttt{noise}. The first seven inputs determine the geometry of simulated CTIS image while the latter three inputs determine the optical parameters.
	
\begin{figure}[htp]
		\begin{center}
			\subfloat[]{\includegraphics[width = 0.49\textwidth]{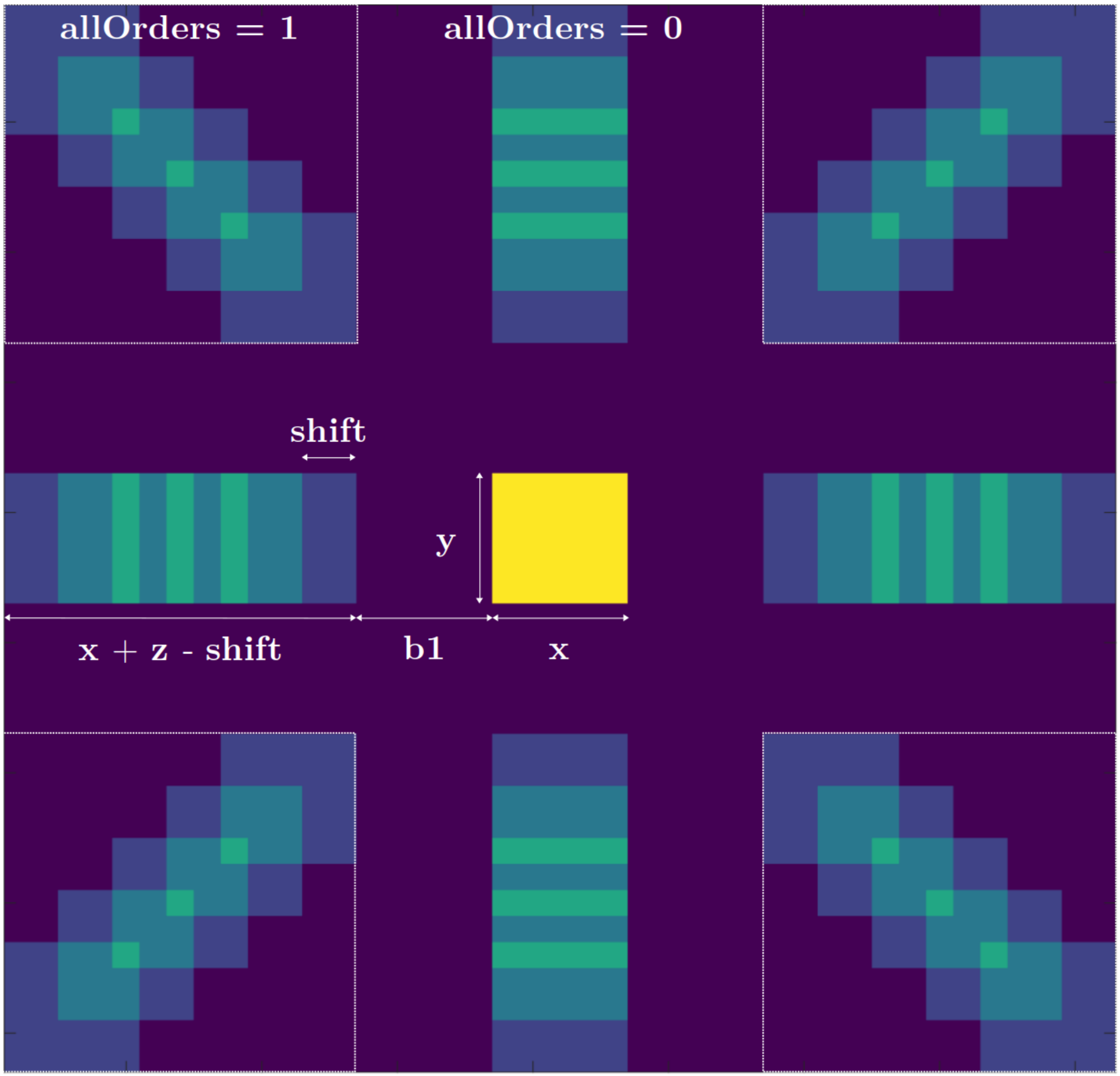}\label{fig:input_a}} \hfill 
			\subfloat[]{\includegraphics[width = 0.469\textwidth]{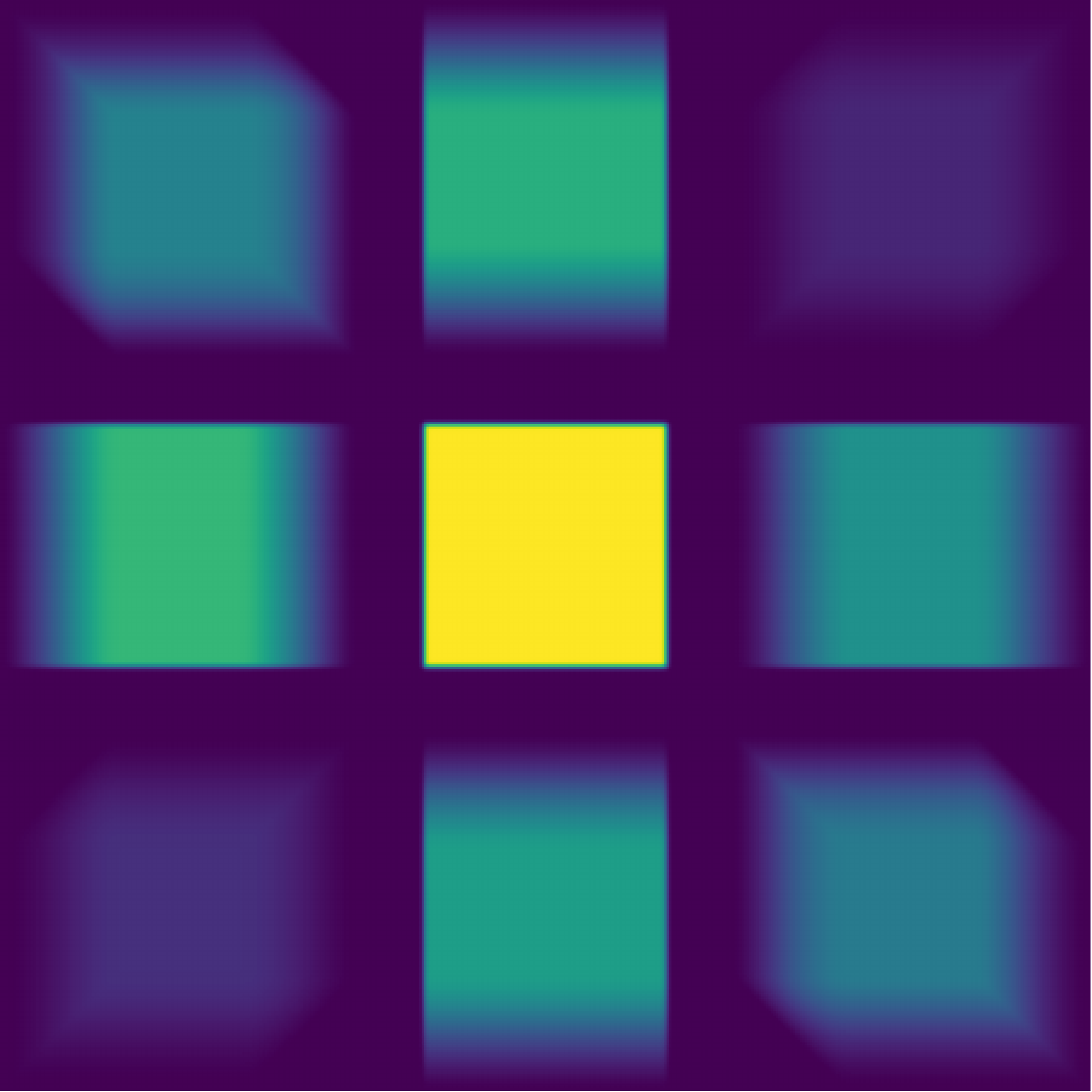}\label{fig:input_b}}
		\end{center}
		\caption{(a) Visual overview of the correspondence between the geometric inputs of the CTIS simulator and the simulated CTIS image for inputs \texttt{x}, \texttt{y}, \texttt{z}, \texttt{b1} = 5, \texttt{b2} = 0, \texttt{shift} = 2, \texttt{allOrders} = true, and without \texttt{sigma\_psf}, \texttt{diff\_sens}, \texttt{illum} or \texttt{noise}. (b) Simulated CTIS image for \texttt{x},\texttt{y} = 100, \texttt{z} = 25, \texttt{b1} = 27, \texttt{b2} = 0, \texttt{shift} = 2, \texttt{allOrders} = true, \texttt{sigma\_psf} = 1.04 and \texttt{noise} = 0.44. \texttt{diff\_sens} and \texttt{illum} are set to the values defined in Section~\ref{app:sensitivity}}
		\label{fig:input_overview}
	\end{figure}

	Figure~\ref{fig:input_a} shows an overview of how the geometric inputs correlate to the simulated CTIS image: \texttt{x}, \texttt{y}, and \texttt{z} denote the two spatial and one spectral dimension of the input respectively, i.e. the zeroth order size is defined by \texttt{x}$\times$\texttt{y} and \texttt{z} denotes the number of spectral channels. \texttt{b1} and \texttt{b2} determine the number of pixels between the zeroth- and first orders and the first orders and the outer border of the image, respectively. \texttt{b2} is usually set to zero, but provided for generalizability. The \texttt{shift} input determines the pixel-shift between consecutive spectral channels in the first order diffraction spots. It should be mentioned, that the user is responsible for setting the inputs to the correct values to match the true CTIS system, i.e. \texttt{b1}, \texttt{b2} and \texttt{shift} are not calculated in the simulator based on the wavelengths of the spectral channels. The \texttt{allOrders} input is a boolean parameter, which simulates all eight surrounding first order diffraction spots if set to true and only the four (no skew orders) first order diffraction spots if set to false. 
	
	The optical parameters of the CTIS system is defined by the \texttt{diff\_sens}, the \texttt{illum} and the applied \texttt{noise}. Additionally, the point-spread-function (PSF), \texttt{sigma\_psf}, is set within the simulator, but is not available to the user as an input. The measurement of the PSF for the true CTIS system is detailed in Section~\ref{app:psf}. The \texttt{diff\_sens} describes the wavelength- and diffraction order dependent sensitivity and must be a matrix of dimensions $\# \text{ orders} \times$\texttt{z}, i.e. in the case of \texttt{allOrders = true}, $9\times$\texttt{z}. The acquisition of the diffraction sensitivity is described in Section~\ref{app:sensitivity}. The illuminant used in the simulator is a standard halogen-tungsten lamp, which is used for the acquisition of true CTIS images. The illumination input is a vector of size $1\times z$, which contains the spectrum of the illuminant measured with an intensity calibrated Avaspec 2048x14 spectrometer operating in the 200-720~nm range. The spectrometer was calibrated using a radiometrically calibrated light source (HL-3P-CAL) from Ocean Insights.
	
	Figure~\ref{fig:input_b} shows a simulated $450\times450$~pixels CTIS image of a $100\times100\times25$ white cube, i.e. all voxels are set to 1, with \texttt{b1} = 27, \texttt{b2} = 0, \texttt{shift} = 2, \texttt{allOrders} = true, \texttt{sigma\_psf} = $\sigma=1.04$~pixels and \texttt{noise} = 0.44. The diffraction sensitivity and illumination is set to the values defined in Section~\ref{app:sensitivity}.
	
	\begin{figure}[htp]
		\begin{center}
			\includegraphics[width = \textwidth]{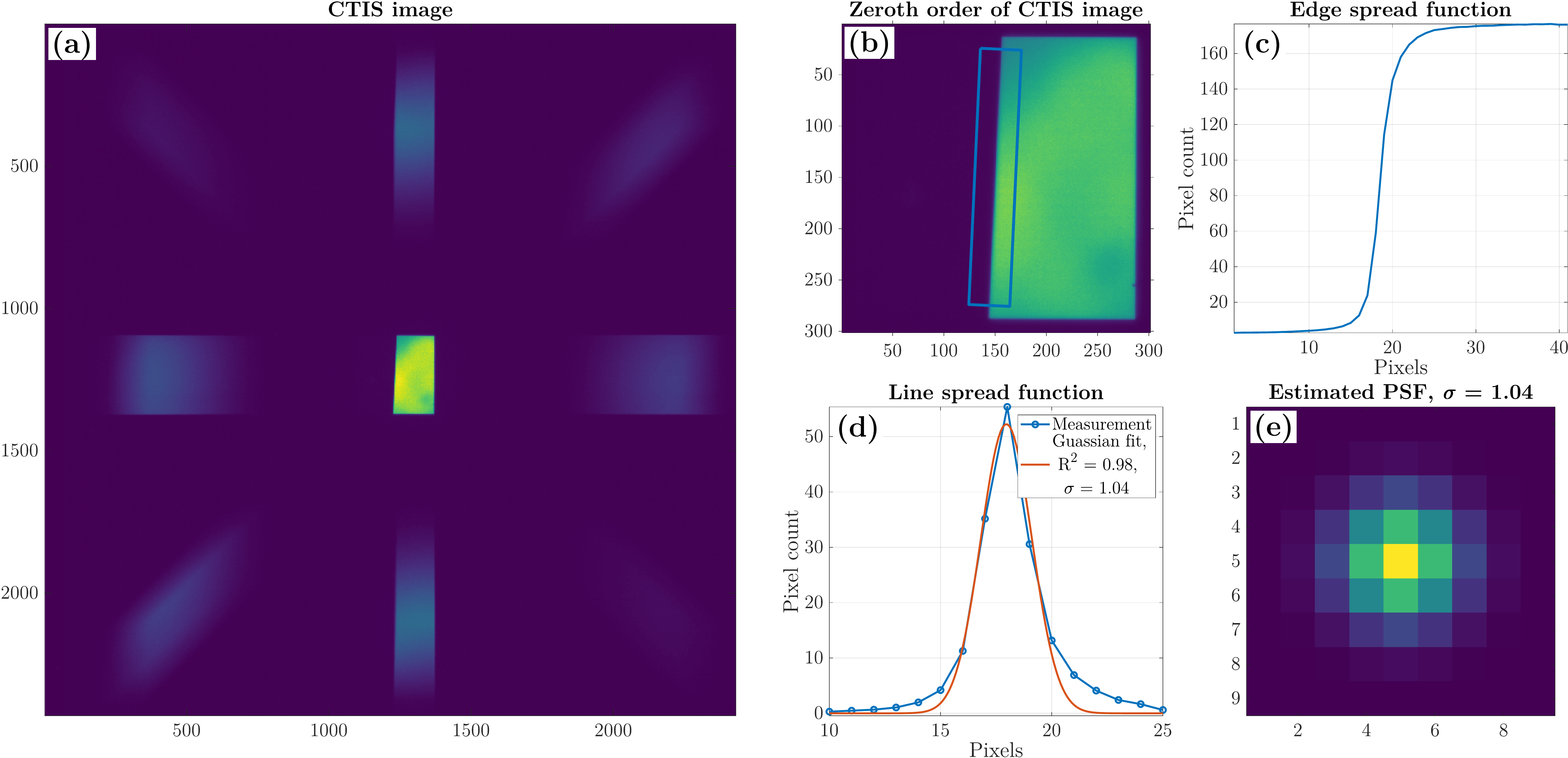}
		\end{center}
		\caption{(a) CTIS image of a slanted razorblade back illuminated by tungsten halogen lamps for point spread function (PSF) estimation. (b) zeroth order of CTIS image with $250\times40$ pixels blue rectangle used to estimate the (c) edge spread function (ESF) (mean across 250 pixels) and (d) line spread function (LSF) (derivative of ESF). The standard deviation of the LSF is estimated from a Gaussian fit, and (e) depicts the estimated symmetric PSF.}
		\label{fig:slanted_edge}
	\end{figure} 
	\section{Measurements of point spread function} \label{app:psf} 
	The PSF of the CTIS system is determined using the slanted-edge method~\cite{estribeau_fast_2004}, where a sharp, slanted image is imaged and the PSF is estimated from the transition of the edge. In practice, a back illuminated razorblade is imaged by the CTIS camera (Figure~\ref{fig:slanted_edge}\red{a}), and the edge spread function (ESF) (Figure~\ref{fig:slanted_edge}\red{c}) is calculated as the mean profile across 250~pixels (blue rectangle) over the transition in the zeroth order (Figure~\ref{fig:slanted_edge}\red{b}). The line spread function (LSF) is depicted in Figure~\ref{fig:slanted_edge}\red{d}, and it is calculated as the derivative of the ESF, which is equivalent to a 2D slice of the PSF. Thus, assuming a spatially symmetric PSF, the LSF is an approximate estimate of the PSF. The standard deviation of the LSF (and PSF) is determined to be 1.04 pixels based on a Gaussian fit with $R^2=0.98$. The effect of the PSF on a single pixel is visualized in Figure~\ref{fig:slanted_edge}\red{e} for a standard deviation of 1.04, which is the value used in the simulator.

	\section{Measurement of diffraction sensitivity}\label{app:sensitivity}
\begin{figure}[htp]
		\begin{center}
			\includegraphics[width = \textwidth]{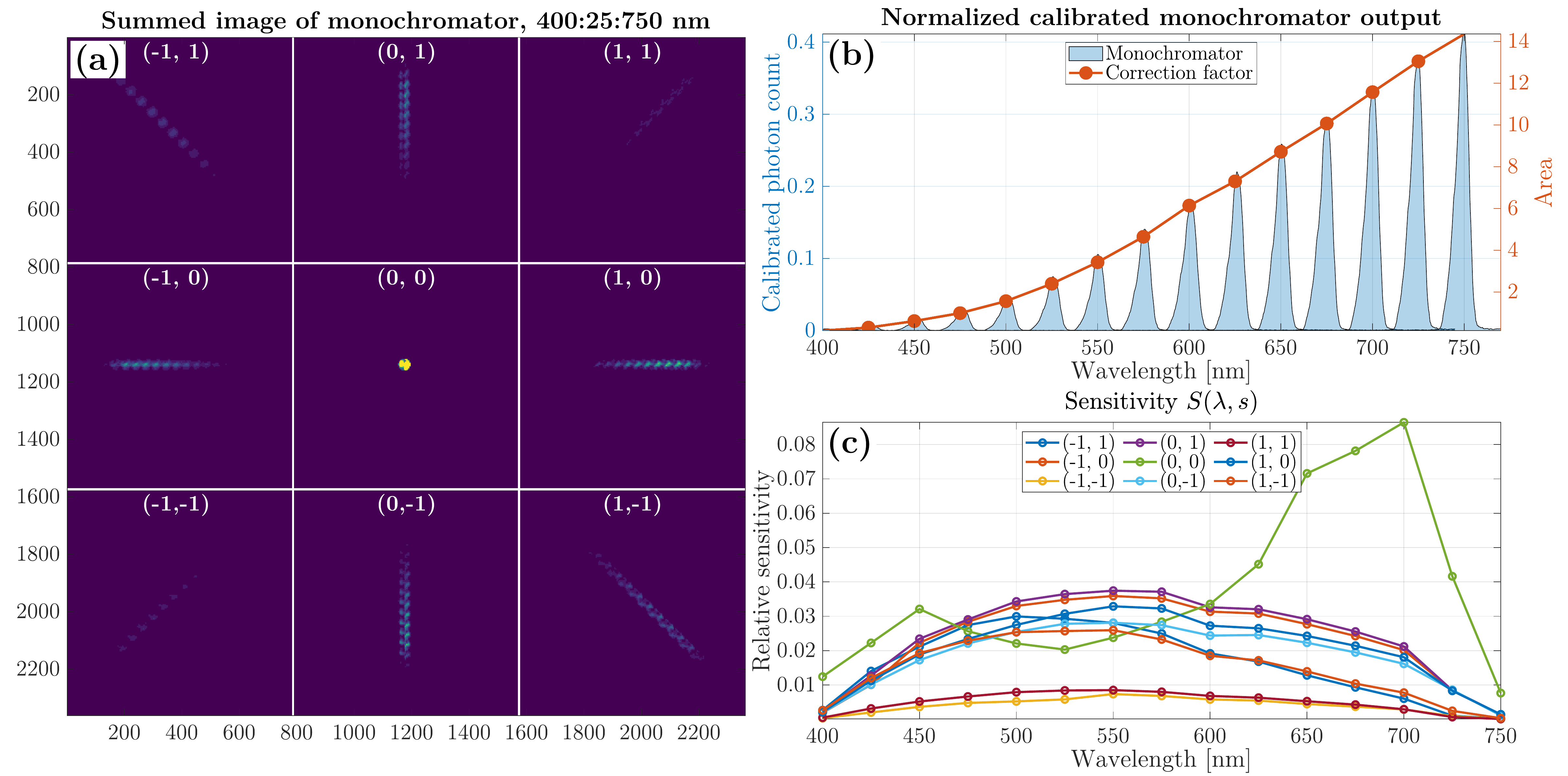}
		\end{center}
		\caption{(a) superimposed image consisting of monochromator images for 400-750 nm with 25 nm steps. (b) Normalized monochromator photon count measured with a calibrated Avaspec 2048x14 spectrometer, and correction factors corresponding to the area under the curve. (c) calculated diffraction sensitivity $S(\lambda,s)$ for the wavelength range 400-750 nm and the 9 diffraction order spots.}
		\label{fig:diff_sens}
	\end{figure}

	The diffraction sensitivity of the CTIS system is measured using a monochromator (Newport mini, model: 78027) equipped with a halogen lamp (Tungsten Halogen Light Source, model:
	78043), which is calibrated with an intensity calibrated Avaspec 2048x14 spectrometer. The spectrometer is calibrated using a radiometrically calibrated light source (HL-3P-CAL) from Ocean Insights. Since the measurements are conducted with the whole CTIS system, both the diffraction efficiency of the DOE, the transmission of the optical system, the CMOS image sensor response and optical aberrations are intrinsically a part of the acquired images. Thus, the measured diffraction sensitivity estimates the combined contribution from all these contributions. The monochromator output is centered in the zeroth order of the CTIS image, and CTIS images and corresponding dark-frame images are captured for wavelengths ranging from 400-750 nm with 25 nm steps. For each wavelength, the exposure time of the CTIS is set to maximize the dynamic range of the image sensor, and the dark frame is captured at the same exposure. Figure~\ref{fig:diff_sens}\red{a} shows the superimposed CTIS images of the monochromator for 400-750 nm.  
	To determine the diffraction sensitivity $S(\lambda,s)$ from the acquired images, the dark frame is subtracted from the CTIS image, and the total photon count $PC(\lambda,s)$ for each diffraction spot (zeroth- and surrounding first orders) is divided with the product of the exposure time $t(\lambda)$ in \textmu s, the camera gain $G(\lambda)$ (set to 1 for all images) and the wavelength correction factor $C(\lambda)$ from the spectrometer:
	\begin{align}
	S(\lambda,s) = \frac{PC(\lambda,s)}{t(\lambda) \cdot G(\lambda) \cdot C(\lambda)},
	\end{align}
	where $\lambda$ and $s$ denote the wavelength in nm and $s$ the diffraction spot, respectively. The monochromator output (Figure~\ref{fig:diff_sens}\red{b}) is used to determine the correction factors $C(\lambda)$, which indicate the relative intensity of the monochromator output (area in blue under the curve). The total photon count of the respective diffraction spots $PC(\lambda,s)$ is determined from the acquired CTIS images as the total volume of each spot. Figure~\ref{fig:diff_sens}\red{c} shows the calculated diffraction sensitivity $S(\lambda,s)$, where the indices ($i$, $j$) of the diffraction spots indicate (rows, columns). The DOE is designed to achieve a symmetric diffraction efficiency in the first orders while minimizing the zeroth order. The diffraction sensitivity of the zeroth order (0,0) is smallest for the lower wavelengths before increasing towards 700 nm, while the first order diffraction sensitivities all follow the same parabolic shape. However, the right-diagonal first orders \{(-1,1) and (1,1)\} have significantly lower diffraction sensitivity across the entire wavelength range. It should be noted, that since $S(\lambda,s)$ is only measured for 15 different wavelengths from 400-750 nm, the sensitivities are interpolated for the 25 and 100 channels used in this work.


	\section{Measurement of halogen lamp illumination}\label{app:illumination}
\begin{figure}[htp]
		\begin{center}´
			\includegraphics[width = \textwidth]{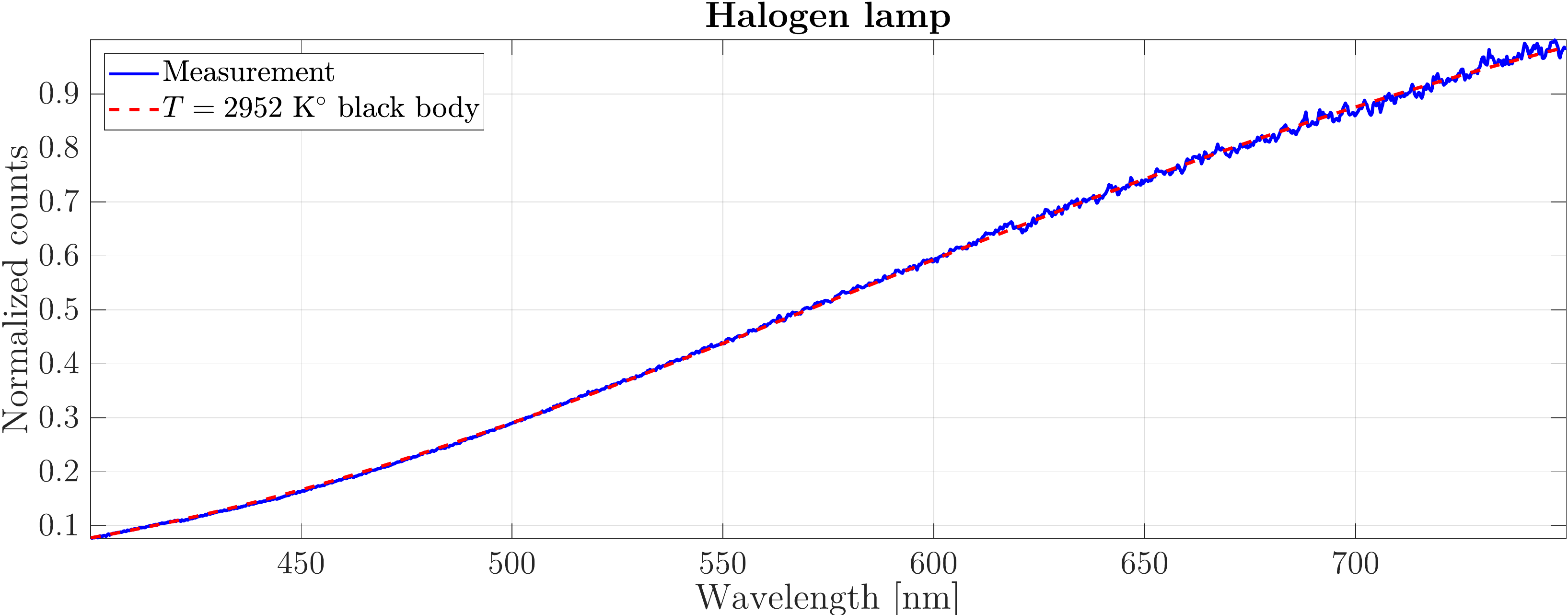}
		\end{center}
		\caption{Measured spectrum of the industrial $3000$~K\degree~halogen lamps used as illumination for CTIS images and a fitted black body with a temperature of $T = 2952$~K\degree.}
		\label{fig:illum}
	\end{figure}

	The intensity calibrated Avaspec 2048x14 spectrometer is used to measure the output of an industrial halogen lamp used as illumination for the acquisition of CTIS images. Figure~\ref{fig:illum} shows the measured spectrum of the halogen lamp and a fitted black body with a temperature of $T = 2952$~K\degree, which is consistent with standard $T = 3000$~K\degree~halogen lamps . 
	
	\section{RGB reconstruction of pushbroom hyperspectral cubes}\label{app:RGB_cubes}
\begin{figure}[h]
		\begin{center}
			\includegraphics[width = \textwidth]{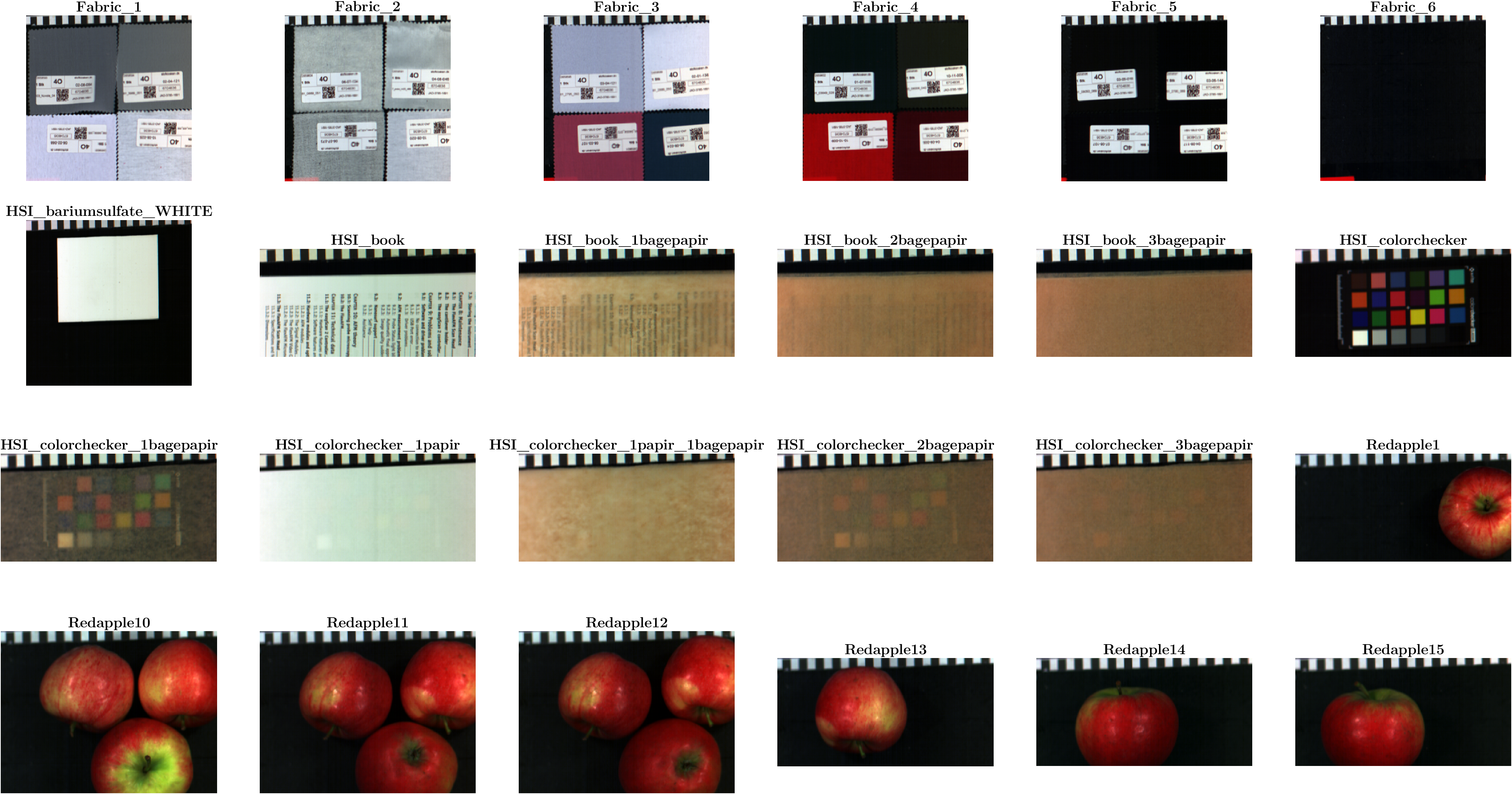}\par\bigskip
			\includegraphics[width = \textwidth]{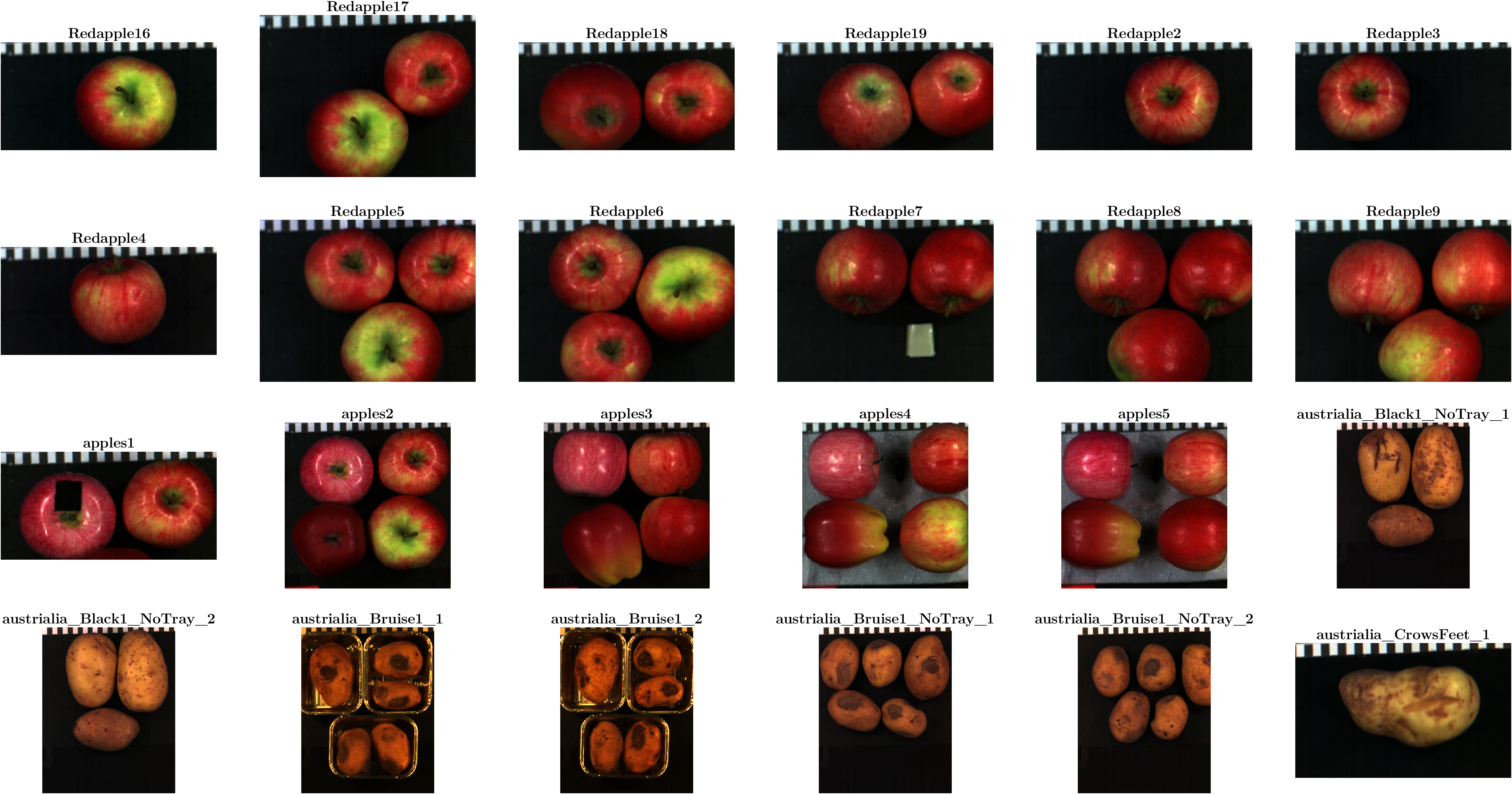}
		\end{center}
		\caption{\textbf{(1)} RGB reconstructions of hyperspectral images captured with the pushbroom system and used in the training of the neural networks.}
		\label{fig:supp_rgb_1}
	\end{figure}

\begin{figure}[h]
		\begin{center}
			\includegraphics[width = \textwidth]{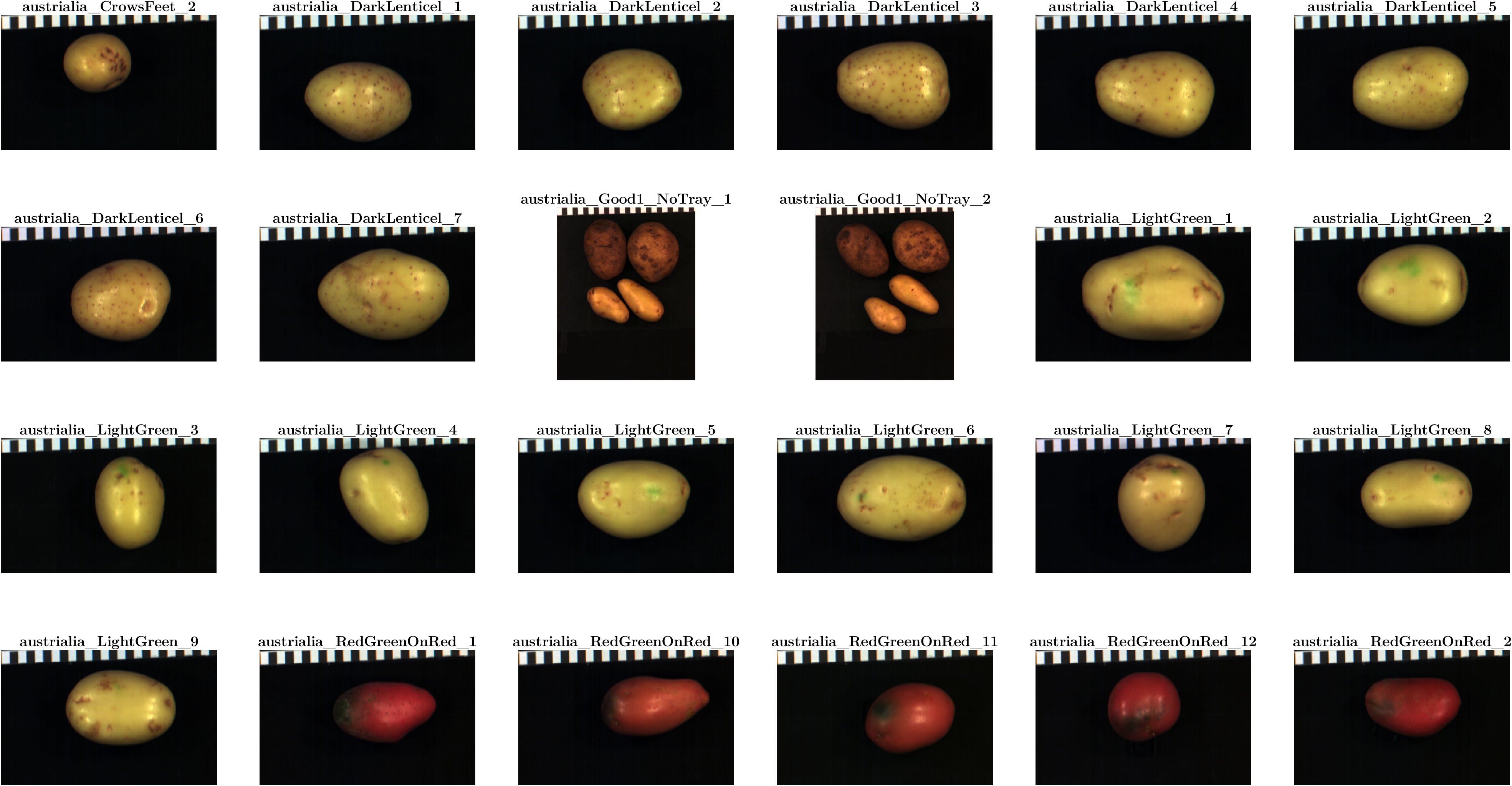}\par\bigskip
			\includegraphics[width = \textwidth]{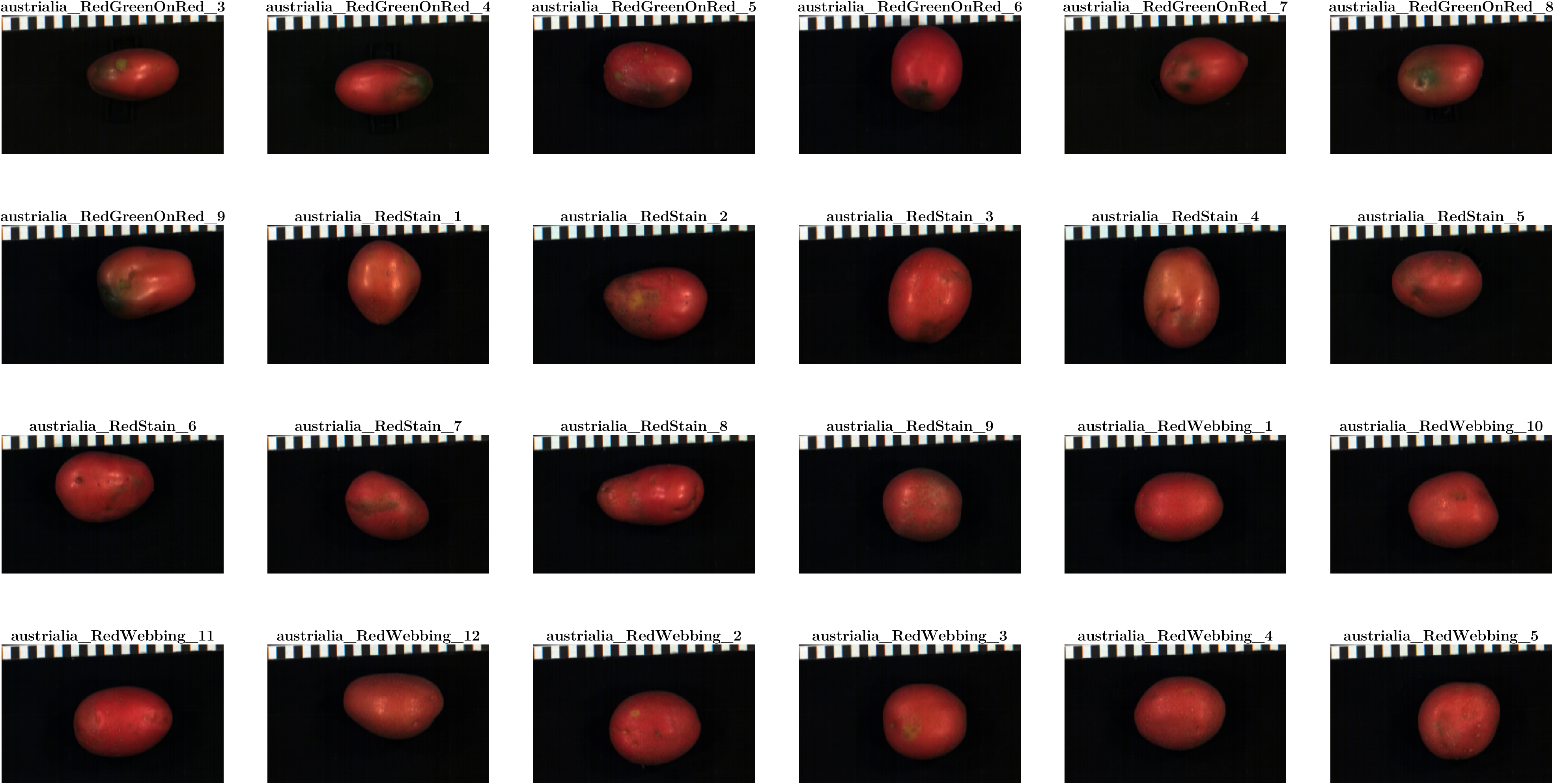}
		\end{center}
		\caption{\textbf{(2)} RGB reconstructions of hyperspectral images captured with the pushbroom system and used in the training of the neural networks.}
		\label{fig:supp_rgb_2}
	\end{figure}

\begin{figure}[h]
		\begin{center}
			\includegraphics[width = \textwidth]{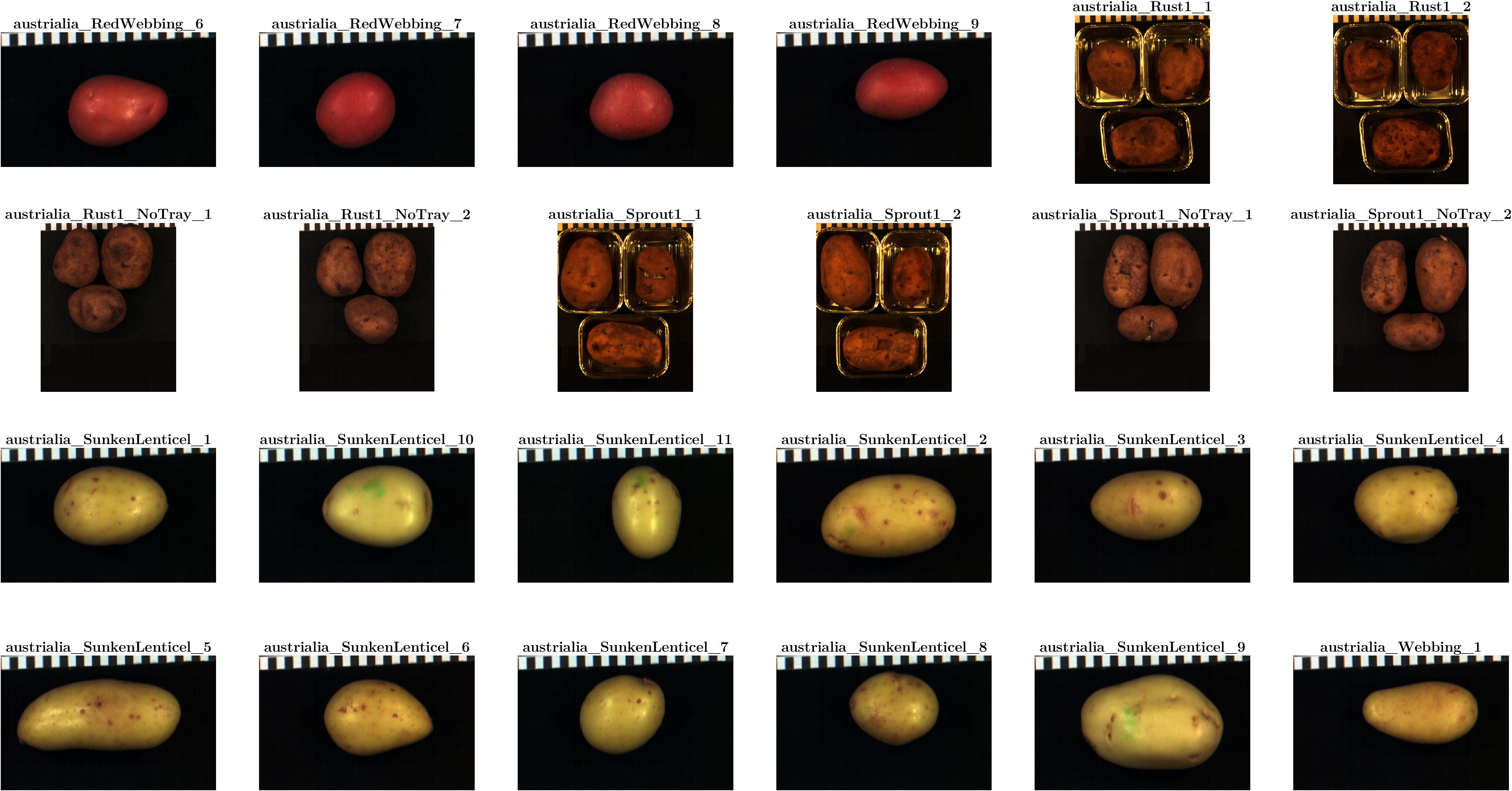}\par\bigskip
			\includegraphics[width = \textwidth]{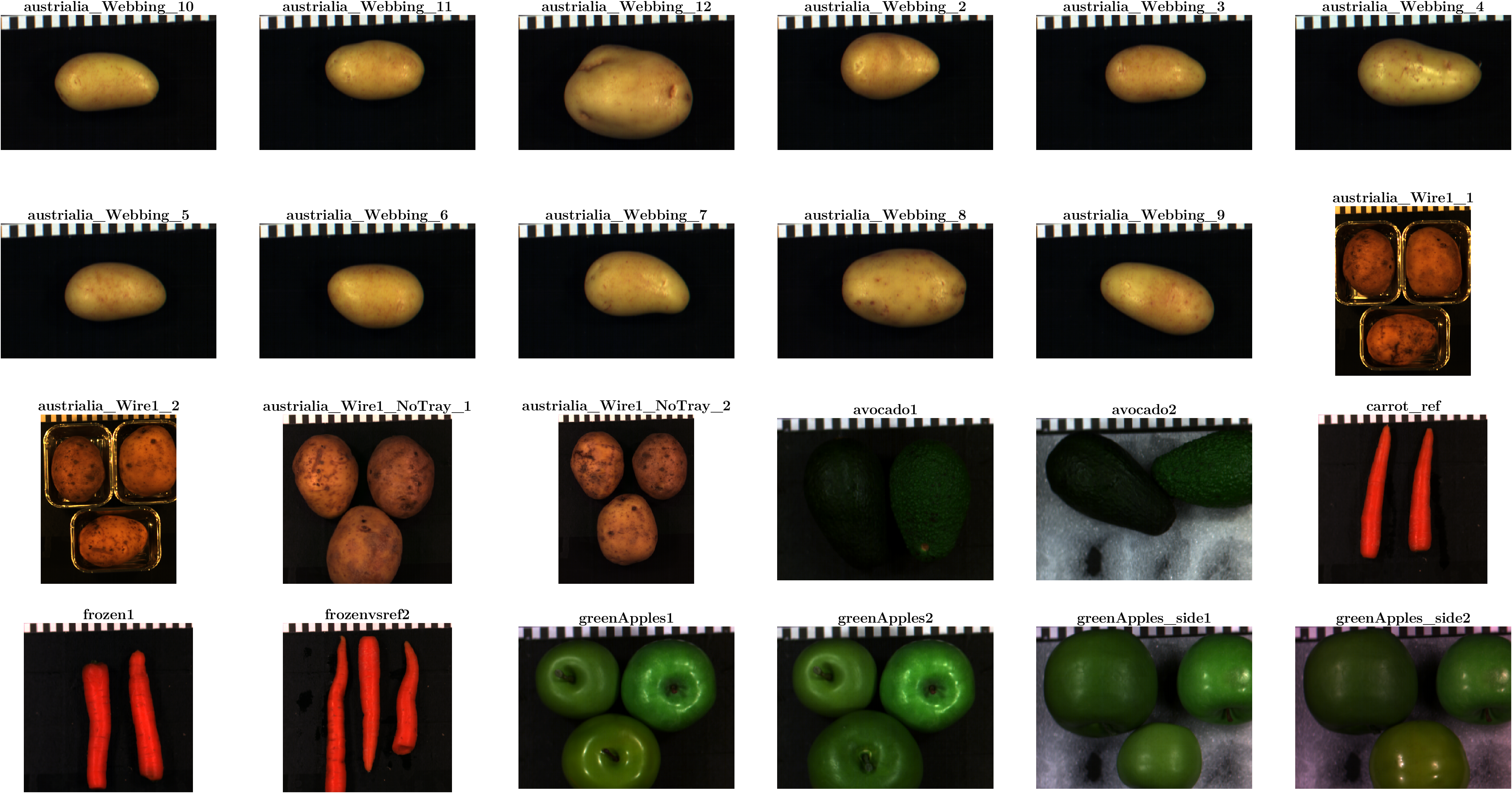}
		\end{center}
		\caption{\textbf{(3)} RGB reconstructions of hyperspectral images captured with the pushbroom system and used in the training of the neural networks.}
		\label{fig:supp_rgb_3}
	\end{figure}

\begin{figure}[h]
		\begin{center}
			\includegraphics[width = \textwidth]{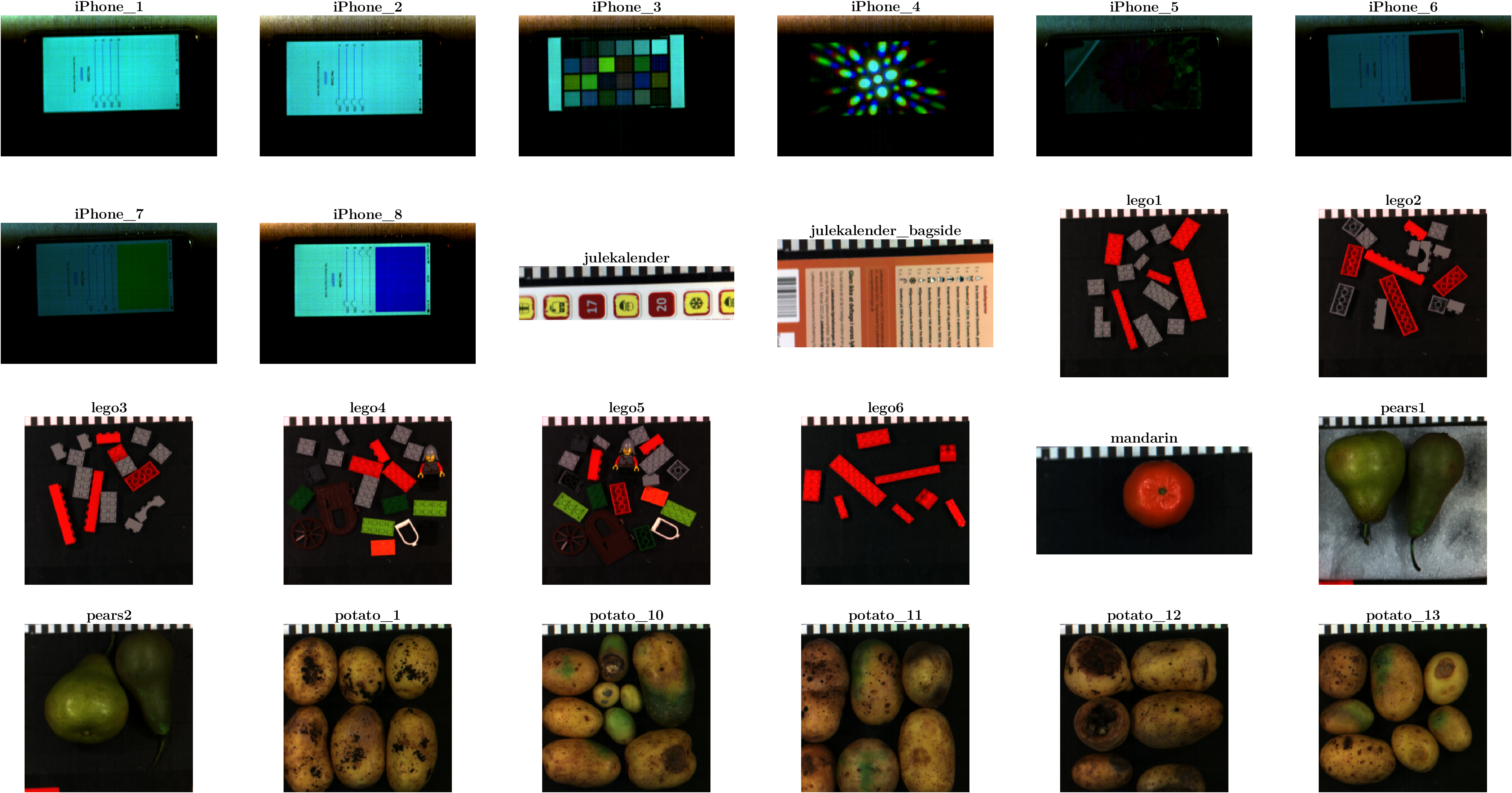}\par\bigskip
			\includegraphics[width = \textwidth]{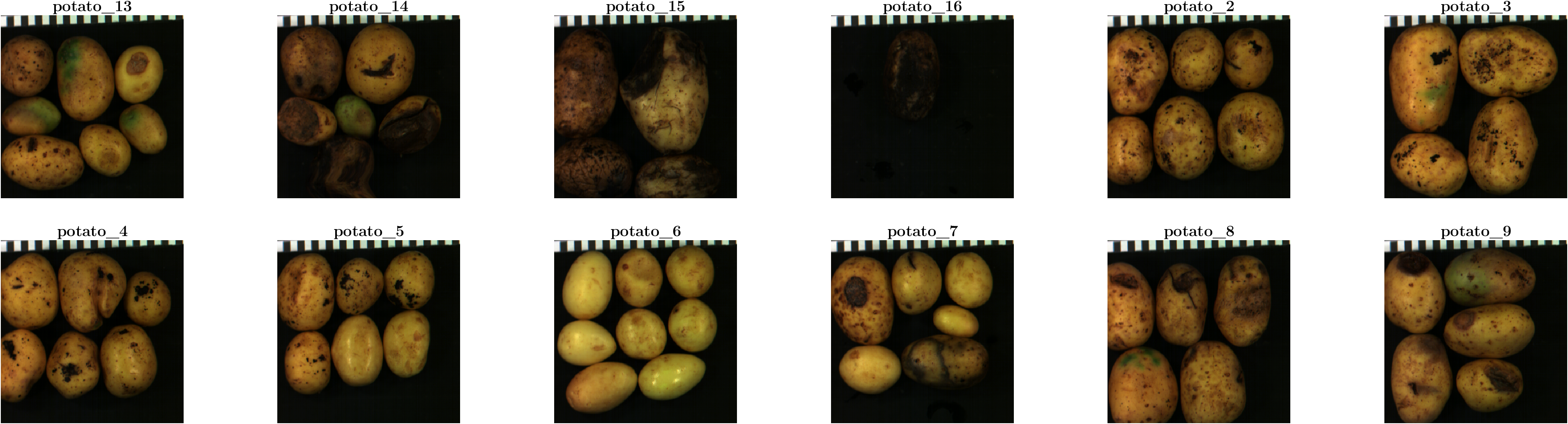}
		\end{center}
		\caption{\textbf{(4)} RGB reconstructions of hyperspectral images captured with the pushbroom system and used in the training of the neural networks.}
		\label{fig:supp_rgb_4}
	\end{figure}

	Figure~\ref{fig:supp_rgb_1}-\ref{fig:supp_rgb_4} shows RGB reconstructions of the 178 hyperspectral images captured with the VIS-NIR pushbroom system, which were used to generate data sets of CTIS images for the networks. The RGB images are generated by combining three spectral channels at 470~nm (blue), 549~nm (green) and 650~nm (red). 
	
	\clearpage
	\section{Comparison of RGB reconstructions for reconstructed hyperspectral cubes - 25 channels}
\begin{figure}[h]
		\begin{center}
			\includegraphics[width = \textwidth]{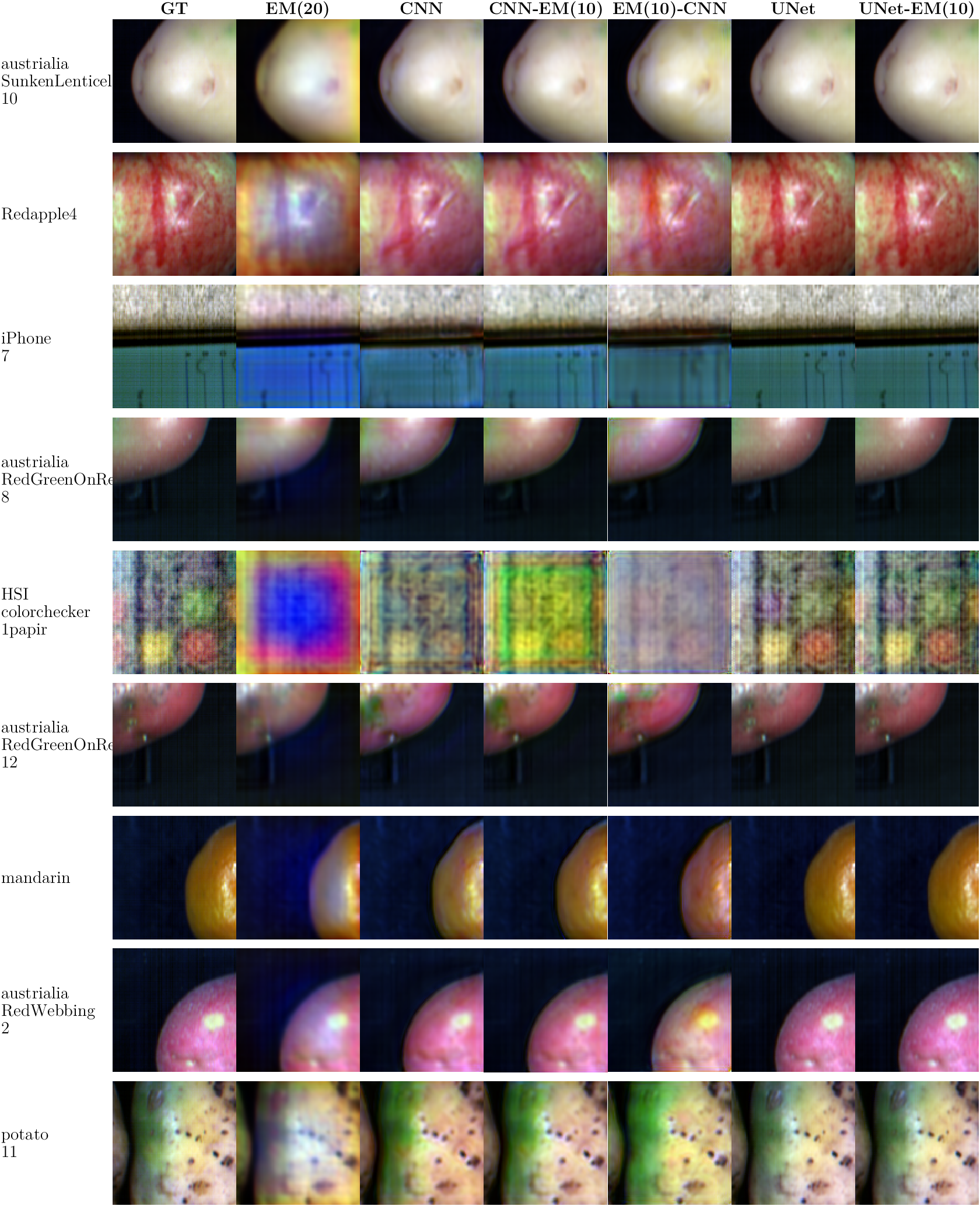}
		\end{center}
		\caption{RGB comparison of reconstructed hyperspectral cubes (seen) for 25 channels. RGB images are generated from spectral channel 7, 9 and 13.}
		\label{fig:recon_rgb_1}
	\end{figure}

\begin{figure}[h]
		\begin{center}
			\includegraphics[width = \textwidth]{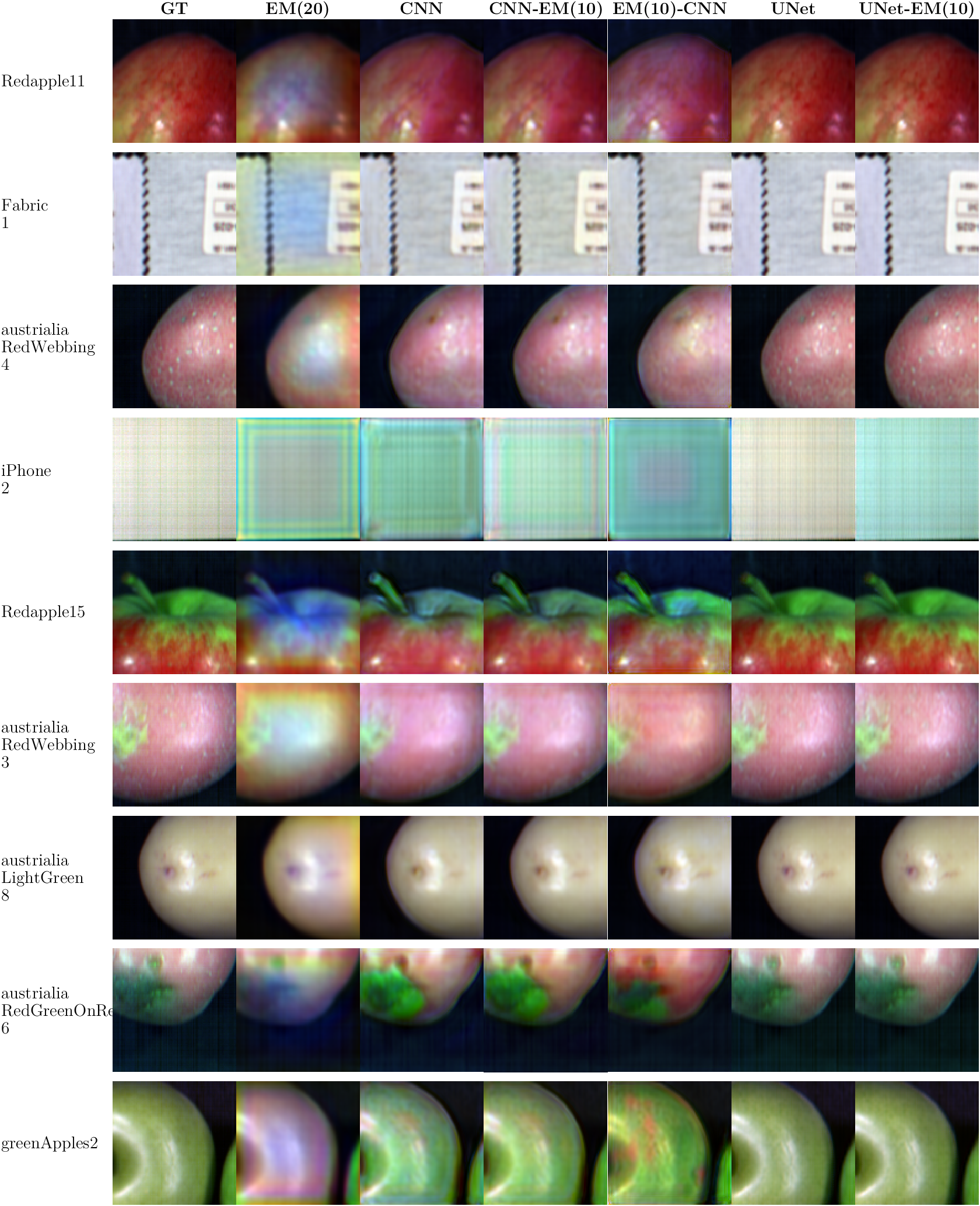}
		\end{center}
		\caption{RGB comparison of reconstructed hyperspectral cubes (seen) for 25 channels. RGB images are generated from spectral channel 7, 9 and 13.}
		\label{fig:recon_rgb_2}
	\end{figure}

\begin{figure}[h]
		\begin{center}
			\includegraphics[width = \textwidth]{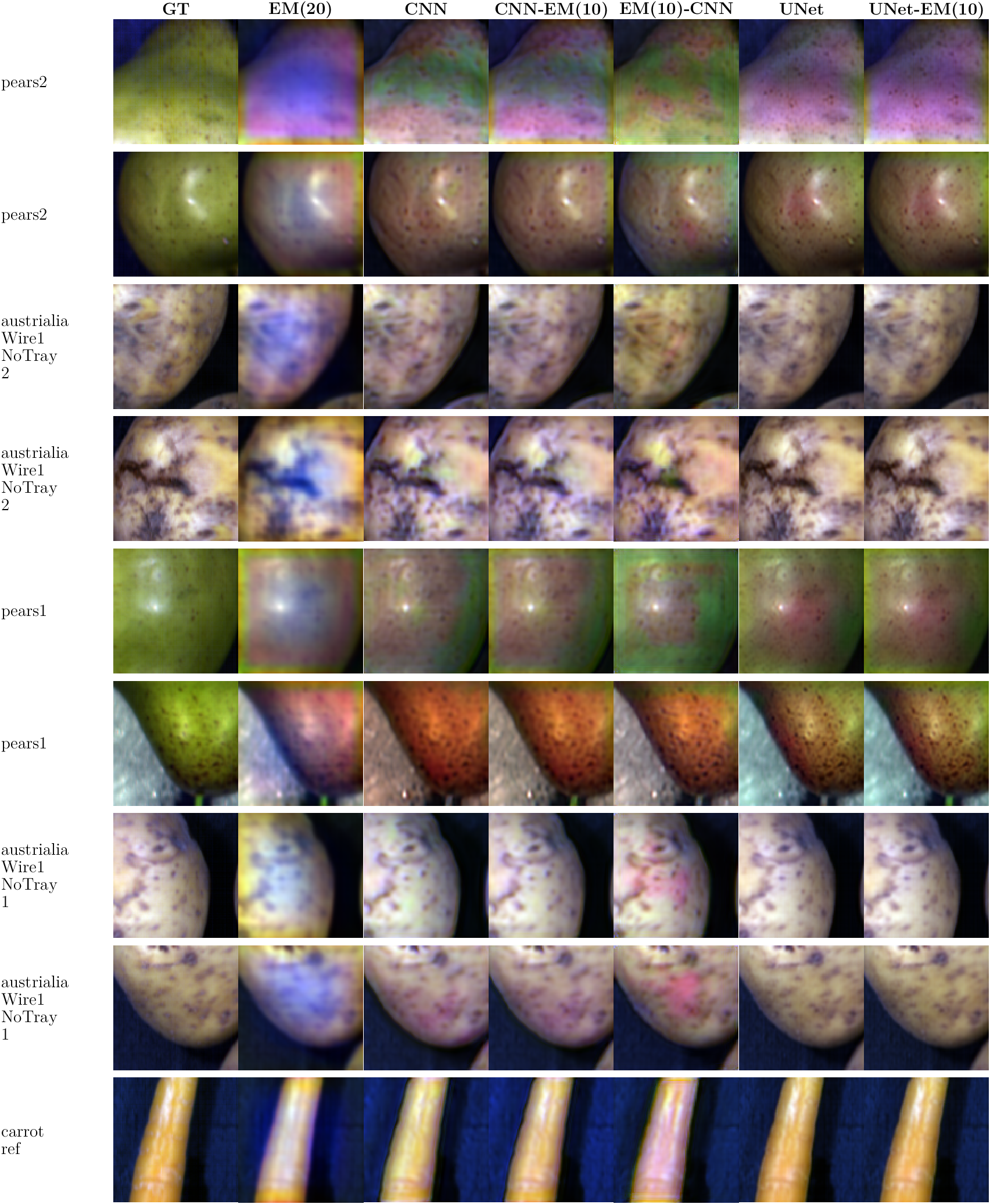}
		\end{center}
		\caption{RGB comparison of reconstructed hyperspectral cubes (unseen) for 25 channels. RGB images are generated from spectral channel 7, 9 and 13.}
		\label{fig:recon_rgb_3}
	\end{figure}

	Figure~\ref{fig:recon_rgb_1}-\ref{fig:recon_rgb_3} show comparisons of the ground truth (GT) RGB visualization with the reconstructed RGB images for EM(20 iterations), CNN, CNN-EM(10 iterations), EM(10 iterations)-CNN, UNet and UNet-EM(10 iterations) for various hyperspectral images. The RGB images are generated by combining the 7th, 9th and 13th channel. Figure~\ref{fig:recon_rgb_1} and \ref{fig:recon_rgb_2} contain seen cubes used in the training, while Figure~\ref{fig:recon_rgb_3} contains unseen cubes.
	
	\newpage
	\section{Comparison of RGB reconstructions for reconstructed hyperspectral cubes - 100 channels}
\begin{figure}[h]
		\begin{center}
			\includegraphics[width = .65\textwidth]{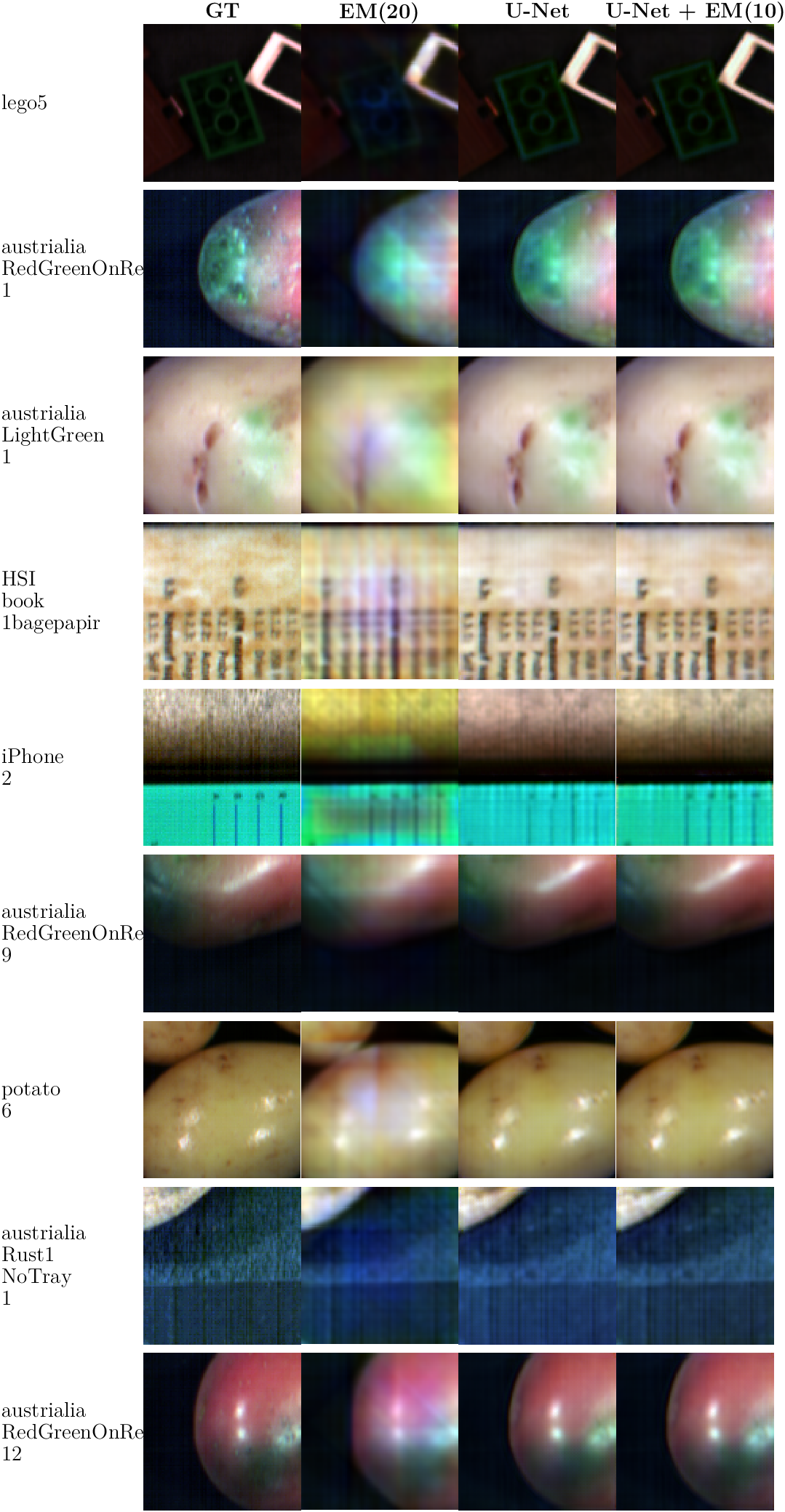}
		\end{center}
		\caption{RGB comparison of reconstructed hyperspectral cubes (seen) for 100 channels. RGB images are generated from spectral channel 14, 29 and 48.}
		\label{fig:recon_100_rgb_1}
	\end{figure}

\begin{figure}[h]
		\begin{center}
			\includegraphics[width = .65\textwidth]{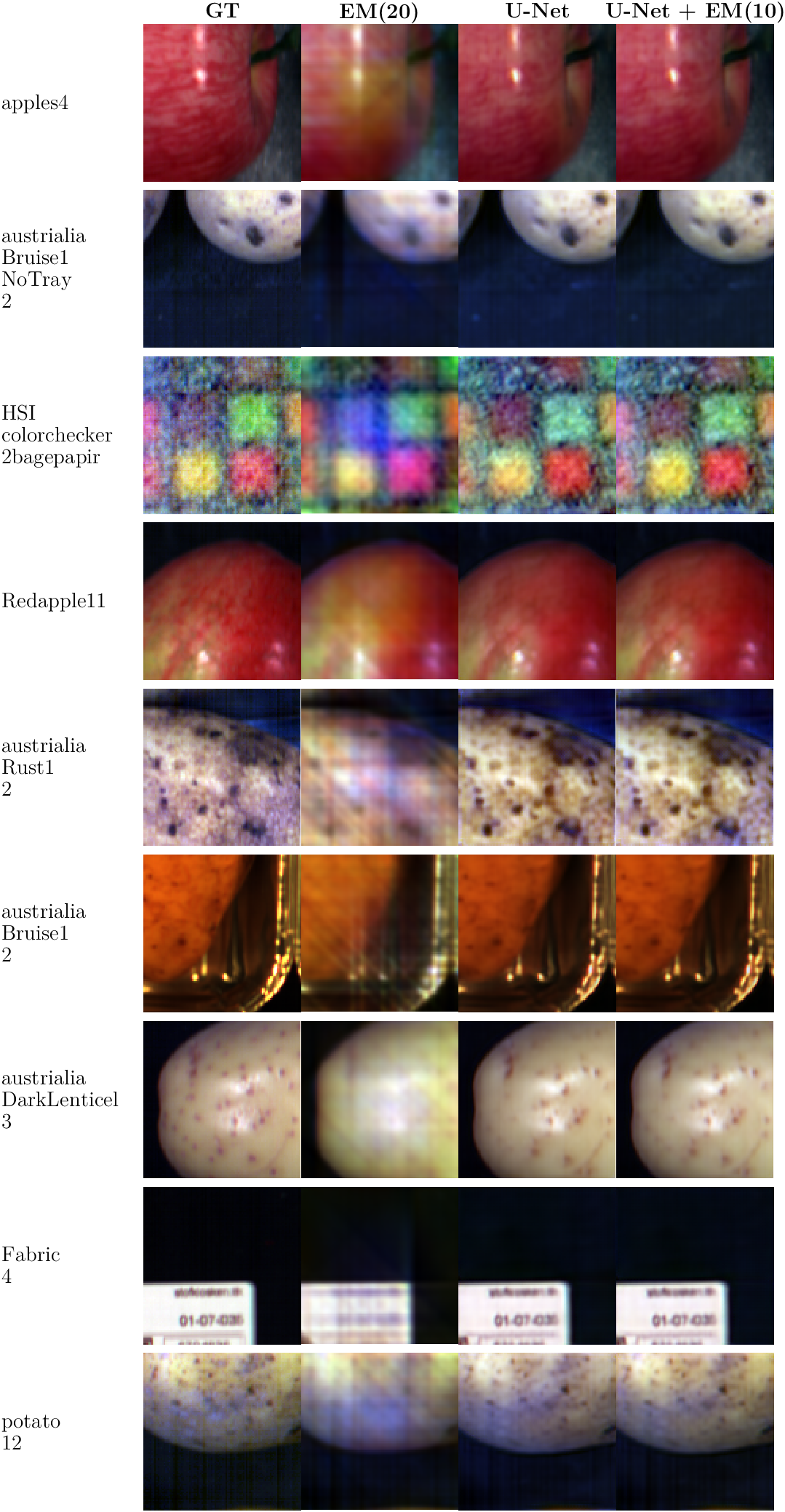}
		\end{center}
		\caption{RGB comparison of reconstructed hyperspectral cubes (seen) for 100 channels. RGB images are generated from spectral channel 14, 29 and 48.}
		\label{fig:recon_100_rgb_2}
	\end{figure}

\begin{figure}[h]
		\begin{center}
			\includegraphics[width = .65\textwidth]{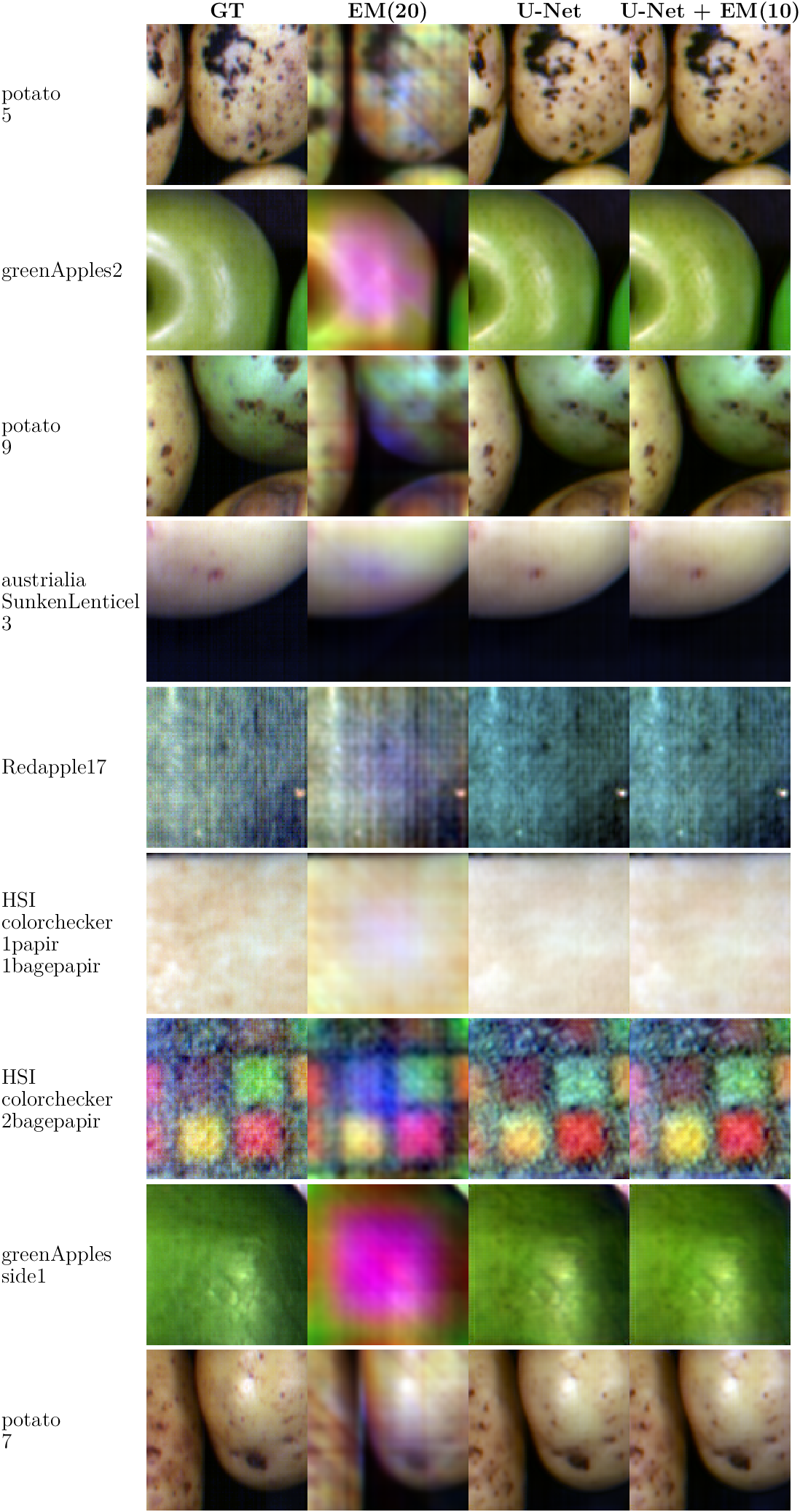}
		\end{center}
		\caption{RGB comparison of reconstructed hyperspectral cubes (unseen) for 100 channels. RGB images are generated from spectral channel 14, 29 and 48.}
		\label{fig:recon_100_rgb_3}
	\end{figure}

	Figure~\ref{fig:recon_100_rgb_1}-\ref{fig:recon_100_rgb_3} show comparisons of the ground truth (GT) RGB visualization with the reconstructed RGB images for EM(20 iterations), UNet and UNet-EM(10 iterations) for various hyperspectral images. The RGB images are generated by combining the 14th, 29th and 48th channel. Figure~\ref{fig:recon_100_rgb_1} and \ref{fig:recon_100_rgb_2} contain seen cubes used in the training, while Figure~\ref{fig:recon_100_rgb_3} contains unseen cubes.
	
	\newpage
	\bibliography{ref_suppl}